\documentclass[12pt]{article}
\usepackage{graphicx} 
\usepackage[letterpaper, margin=1in]{geometry}
\usepackage{amsmath}
\usepackage{authblk}
\usepackage{ulem}
\usepackage{float}
\usepackage{xcolor}
\usepackage{todonotes}
\definecolor{circleblue}{HTML}{4D7FDD}
\DeclareRobustCommand{\filledcircle}{%
  \raisebox{0.05ex}{\textcolor{circleblue}{\scalebox{1.4}{\textbullet}}}%
}

\DeclareRobustCommand{\opencircle}{%
  \raisebox{0.05ex}{\textcolor{circleblue}{\scalebox{1.4}{$\circ$}}}%
}
\usepackage{multirow}
\usepackage{makecell}
\usepackage{array}
\usepackage{pifont}
\usepackage{booktabs}
\usepackage{longtable}
\usepackage[table]{xcolor}
\usepackage{tabularx}
\makeatletter
\usepackage{romannum}
\usepackage[most]{tcolorbox}
\newcounter{boxeq}
\newcommand{\boxeqtag}{%
  \refstepcounter{boxeq}%
  \tag{B\theboxeq}%
}
\makeatother
\usepackage[labelfont=bf]{caption}
\usepackage[numbers,sort&compress]{natbib}
\title{Materials and Metrology for Out-of-Plane Damping-Like Spin-Orbit Torque\footnote{ E-mail: dcr14@cornell.edu}}
\author[1]{Xiaoxi Huang}
\author[1]{Yuhan Liang}
\author[1]{Orion Smedley}
\author[1,2,$\ast$]{Daniel C. Ralph}
\date{}

\affil[1]{Department of Physics, Cornell University, Ithaca, NY 14853, USA}
\affil[2]{Kavli Institute at Cornell, Cornell University, Ithaca, NY 14853, USA}

\begin{document}
\pagenumbering{arabic}
\maketitle

\textbf{
A new generation of energy-efficient, high-density, and non-volatile memory devices could be enabled if a more efficient mechanism is developed for switching the magnetization orientation of nanoscale magnetic devices possessing perpendicular magnetic anisotropy. A promising strategy under widespread investigation is current-induced damping-like spin-orbit torque with a strong unconventional out-of-plane component, rather than only the conventional in-plane component. In recent years, such torques have been reported in a broad range of materials platforms, arising primarily from three classes of symmetry breaking within the material generating the torque: low crystal symmetry in single-crystalline materials, antiferromagnetic order, and ferromagnetic order. Despite this progress, the field remains fragmented. Measurements using different methods often disagree, artifacts have been misinterpreted, and qualitative observations are sometimes overinterpreted as evidence of technological promise.
This Perspective will provide a critical overview of out-of-plane damping-like spin-orbit torques, analyzing the classes of materials which can generate such torques, the experimental methods by which the torques can be quantified, and the progress toward achieving torque efficiencies sufficient for scalable magnetic technologies.
\\
}

\noindent \textbf{\Romannum{1}. Introduction}\\

\begin{figure}[t!]
    	\centering
    	\includegraphics[width=1\textwidth]{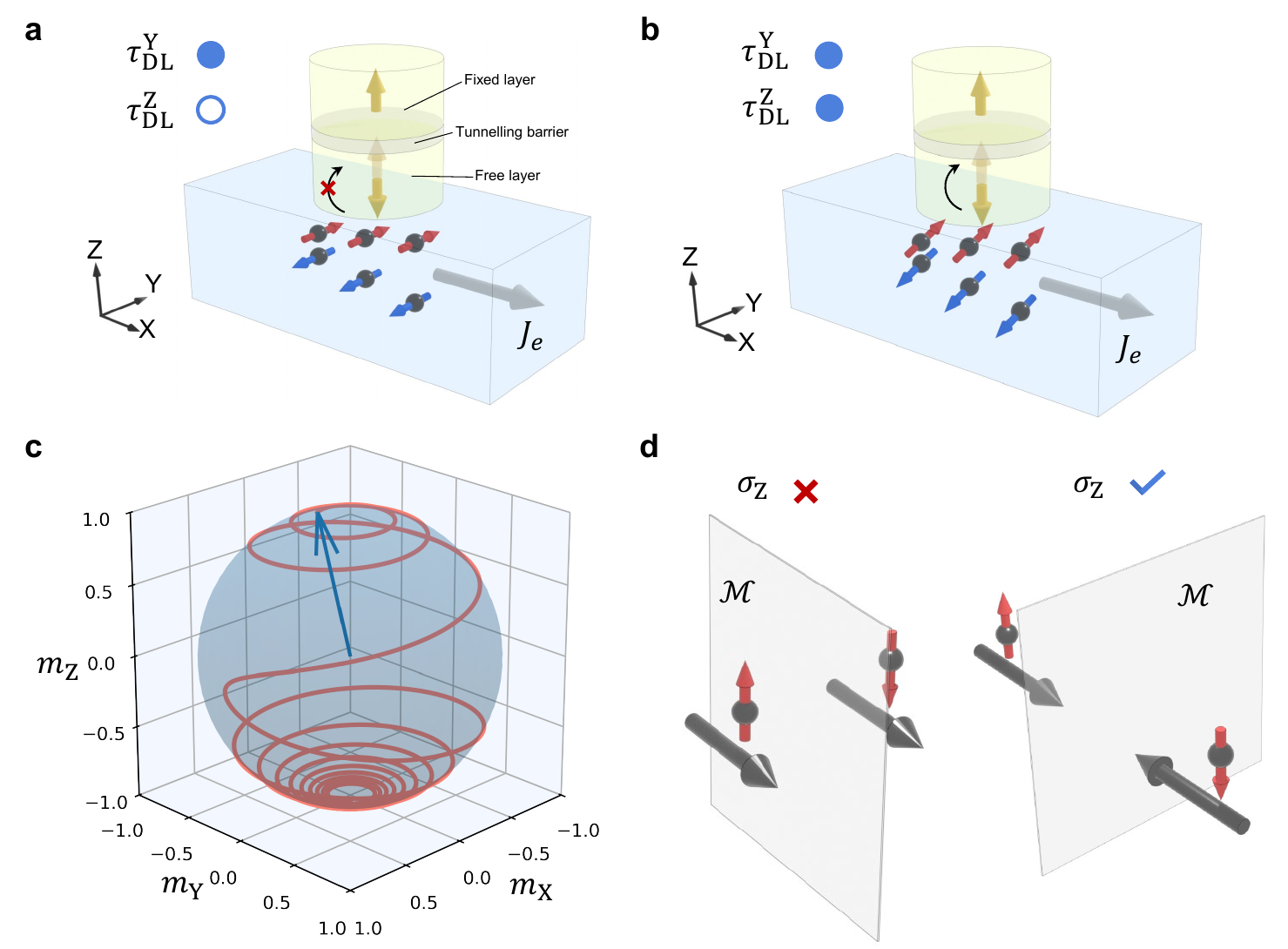}
    	\caption{\textbf{Anti-damping magnetization switching and symmetry constraints of an out-of-plane damping-like torque.} \textbf{a},\textbf{b},
        Anti-damping spin-torque-driven magnetization switching of a PMA magnet is not viable when the damping-like torque has only an in-plane component, $\tau_{\rm DL}^\text{Y}$ (\textbf{a}), but becomes viable when an out-of-plane damping-like torque component, $\tau_{\rm DL}^\text{Z}$, is present (\textbf{b}). Arrows in gray indicate the applied charge current, $J_{\rm e}$. \filledcircle{} and \opencircle{} indicate the presence and absence of the corresponding components of the spin-orbit torques, respectively. \textbf{c,  }Representative trajectory of anti-damping switching in a PMA magnet, shown for switching from the down state to the up state. 
        \textbf{d, }Schematic illustrating one of the symmetry constraints for generating an out-of-plane spin-moment component, $\sigma_\text{Z}$. $\mathcal{M}$ denotes a mirror plane.
       }
    	\label{fig:1}
    \end{figure}

Damping-like spin-orbit torque (SOT) exerted by spin currents produced in materials with strong spin-orbit coupling\cite{mihai2010current,liu2011spin,miron2011perpendicular,liu2012spin,nan2019anisotropic,huang2021novel,liang2023field,huang2024manipulating}, topological band structures\cite{mellnik2014spin,fan2014magnetization,fan2016electric,han2017room,dc2018room}, or magnetic order\cite{kimata2019magnetic,nan2020controlling,chen2021observation,huang2026turn} is a leading candidate to enable efficient switching of magnetization for magnetic memory devices\footnote{We will utilize the term ``spin-orbit torque'' loosely, applying it to any case in which an applied electric field generates a perpendicularly-flowing spin current to exert a torque on a magnetic moment, even if the microscopic mechanism of the spin-current generation is not dominated by spin-orbit interactions.}.  To realize thermally-stable nanoscale memory elements, magnetic random-access memory (MRAM) must be based on magnets with perpendicular magnetic anisotropy (PMA)\cite{worledge2024spin}. Unfortunately, common spin-source materials with high crystal symmetry cannot drive efficient SOT switching of nanoscale PMA magnets.  This is because ordinary high-symmetry spin-source materials are constrained by symmetry; they can only produce spin polarizations and associated damping-like torques oriented within the sample plane, and this orientation cannot drive switching of nanoscale PMA magnets by the most efficient anti-damping mechanism (Fig.~\ref{fig:1}a).  To achieve anti-damping switching of nanoscale magnets with PMA requires that the damping-like torque have a strong out-of-plane component (Fig.~\ref{fig:1}b, c). \\

The literature has sometimes been unclear about this point.  It is true that a conventional in-plane damping-like torque can drive switching reasonably efficiently via domain wall motion in  much-larger-than-nanoscale PMA magnetic devices, as long as a symmetry-breaking magnetic field is applied\cite{miron2011perpendicular,emori2013current,lee2014central}. This has led to significant research into different ways of applying effective fields to enable the SOT switching in micron-scale devices to become ``field-free''.  However, the domain-wall mechanism is not available for practical memory elements approaching the 10 nm scale, because they are too small to host domain walls.  In the absence of domain walls, switching of nanoscale PMA devices by conventional in-plane SOT, even with a symmetry-breaking field, is inefficient, requiring very large current densities approximately 30 times greater than for micron-scale samples\cite{zhang2015spin}. The difficulty here is that in nanoscale PMA samples the SOT must directly overcome the energy barrier associated with the anisotropy field (i.e., with a torque $\tau_{\rm DL}^{\rm Y} \sim \gamma B_\text{anis}$, where $\gamma$ is the gyromagnetic ratio and $B_\text{anis}$ is the anisotropy field). As an alternative, an out-of-plane component of damping-like torque has the potential to drive switching with a threshold corresponding to compensating the magnetic damping rather than directly overcoming an energy barrier.  This merely requires a torque of order $\tau_{\rm DL}^{\rm Z} \sim \alpha \gamma B_\text{anis}$\cite{ralph2008spin}, where $\alpha$ is the Gilbert damping constant, often of order 0.01. In addition, this type of efficient anti-damping switching by its nature requires no separate symmetry-breaking magnetic field. We will go into more detail below about the distinction between domain-wall motion and true anti-damping magnetic reversal.\\

To enable an out-of-plane damping-like torque that can enable true anti-damping switching of nanoscale PMA magnets, the symmetry of the spin-source material must be reduced. A detailed symmetry analysis is given by refs.\cite{wimmer2016,manchon2019current}. Fig.\ref{fig:1}d shows one example of how a symmetry can constrain the orientation of SOT;  if a sample has a mirror plane perpendicular to the sample plane, then an out-of-plane damping-like torque is forbidden for charge current flow parallel to the mirror plane. A two-fold rotational symmetry about an axis normal to the sample plane also requires that the out-of-plane damping-like torque be zero.\\

This Perspective seeks to summarize the status of current experiments. After a background discussion with a definition of terms, we will first define the requirements for a useful technology. We will then discuss two main approaches for measuring SOTs: spin-torque ferromagnetic resonance (ST-FMR) and the second-harmonic Hall technique, both of which are capable of quantifying both conventional in-plane damping-like torques and also unconventional torque components.  We will describe the strengths and weaknesses of the two approaches, and what steps should be taken to accomplish accurate measurements.  We will then review the progress toward discovering materials that can generate out-of-plane damping-like torques that are efficient enough to be useful for magnetic technologies, covering the three main types of symmetry breaking within spin-source materials that are under investigation:
(a) single-crystalline materials with low crystal symmetry\cite{macneill2017control,dc2023observation,liu2023field,zhang2023room,bainsla2024large,wang2024field},
(b) antiferromagnetic spin order, either non-collinear\cite{nan2020controlling,cao2023anomalous,nakatsuji2015large,rimmler2025non,han2025unconventional} or collinear\cite{bose2022tilted,karube2022observation,bai2025all,vsmejkal2020crystal,gonzalez2021efficient,feng2022anomalous,krempasky2024altermagnetic}, and
(c) ferromagnetic order in which the torque orientation  is controllable by changing the magnetization direction of the source layer\cite{dyakonov2009,taniguchi2015spin,manchon2015,humphries2017observation,baek2018spin,iihama2018spin,seki2019large,safranski2019spin,safranski2020planar,ryu2022,yang2024field,montoya2025anomalous,huang2025control}.

\noindent \textbf{\Romannum{2}. Background with definition of terms}\\

Both ST-FMR and second-harmonic Hall measurements operate by applying a known charge current to a spin source/ferromagnet bilayer sample to generate the SOT, and then measuring resistance changes associated with the magnetic deflection caused by the  torque. We will call the magnet to which the torque is applied the detecting ferromagnet or the free-layer ferromagnet. \\
 
In the standard analysis approach for both measurement strategies, the magnitude of the magnetization in the free magnetic layer is assumed to remain constant, independent of the applied current\footnote{Note that this assumption may fail, as discussed later in this Perspective}. The motion of the sample magnetization orientation $\hat{m}$ is then modeled by the Landau-Lifshitz-Gilbert-Slonczewski equation including SOTs\cite{slonczewski1996current,liu2011spin,karimeddiny2020transverse}:
\begin{equation}
    \frac{d\hat{m}}{dt}=\gamma\hat{m}\times \frac{dF}{d\hat{m}} + \alpha \hat{m}\times \frac{d\hat{m}}{dt}+
    \vec{\tau}_\text{CI}.
\end{equation}
Here $F$ is the free energy density of the ferromagnet, $\gamma=2\mu_\text{B}/\hbar$ is the gyromagnetic ratio, with $\mu_\text{B}$ being the Bohr magneton, $\alpha$ is the effective Gilbert damping parameter, and $\vec{\tau}_\text{CI}$ is the current-induced torque per unit magnetization.  (We use a convention that $\vec{\tau}_\text{CI}$ has frequency units.)
For analyzing the magnetic dynamics we will use the axes X (parallel to the applied charge current in the sample plane), Y (perpendicular to the charge current in the sample plane), and Z (perpendicular to the sample plane), and describe the magnetization orientation in terms of spherical polar coordinates $\theta$ (polar angle measured relative to the Z axis) and $\varphi$ (azimuthal angle measured relative to the X axis). \\

The current-induced torque $\vec{\tau}_\text{CI}$ can be expressed as the sum of a magnetization-even damping-like torque $\vec{\tau}_\text{DL}$ ($\propto \hat{m}\times(\hat{\sigma} \times \hat{m})$ to first order \cite{garello2013symmetry}) and a magnetization-odd field-like torque $\vec{\tau}_\text{FL}$ ($\propto\hat{\sigma} \times \hat{m}$ to first order).

\begin{figure}[t!]
    	\centering
    	\includegraphics[width=1\textwidth]{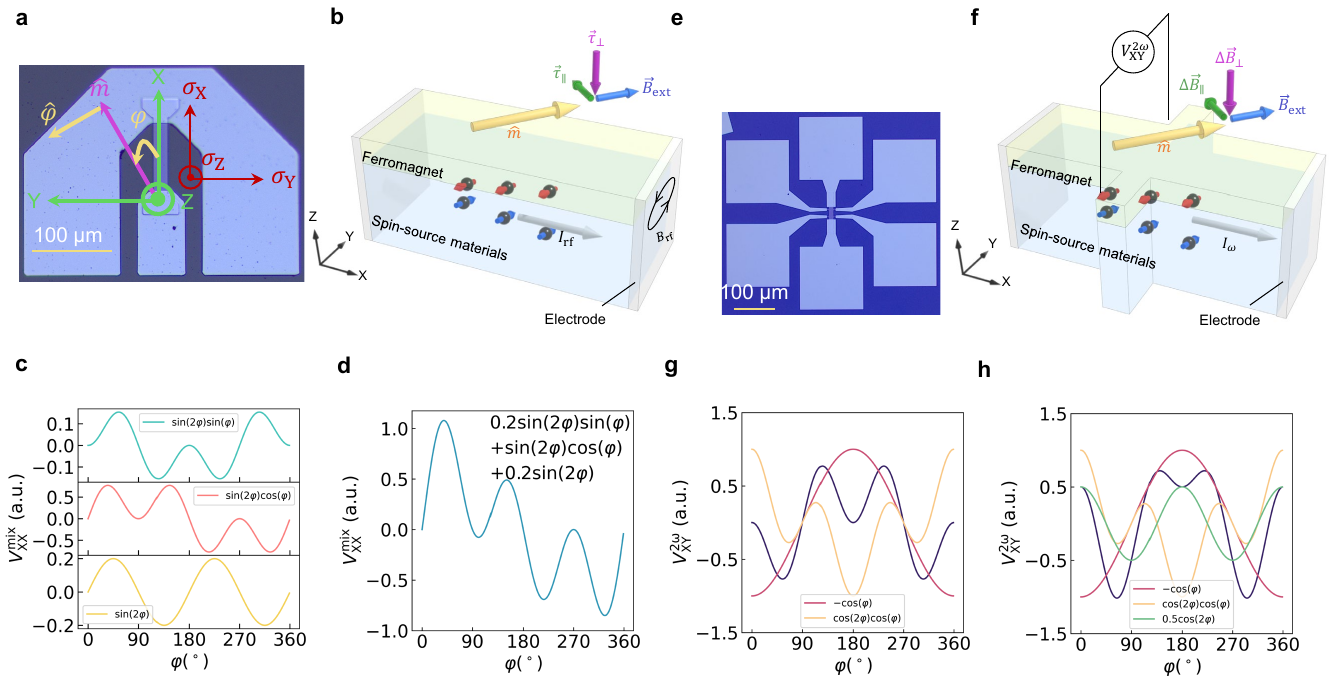}
    	\caption{\textbf{Illustration of two experimental techniques for quantifying out-of-plane damping-like torques.} \textbf{a,} Optical image of a ST-FMR device consisting of ground-signal-ground contacts to a channel containing the spin source/magnet bilayer sample under test. \textbf{b, }Illustration of the ST-FMR mechanism. $\vec\tau_\parallel$ and $\vec\tau_\perp$ are the vector components of the spin-orbit torque in $\hat{\varphi}$ and $\hat{\theta}$ direction. $\vec B_\text{ext}$ is the applied external magnetic field. $I_{\rm rf}$ is the applied microwave current within the device. $B_\text{rf}$ is the current-induced Oersted field.  \textbf{c,} Magnetic-field angle $\varphi$ (deg) dependences of the mixing voltages (arbitrary units) associated with the X, Y, and Z components of SOT. \textbf{d, }Example data for the ST-FMR technique  of mixing voltage as a function of magnetic-field angle contributed by the combined effects of X ($\sin(2\varphi)\sin(\varphi)$), Y ($\sin(2\varphi)\cos(\varphi)$) and Z ($\sin(2\varphi)$) components of damping-like SOT. \textbf{e, }Optical image of a Hall device for second-harmonic Hall voltage measurement. \textbf{f,} An illustration of the second harmonic Hall voltage mechanism. $I_{\omega}$ is the applied a.c. current with a frequency of $\omega$. $\Delta \vec B_\parallel$ and $\Delta \vec B_\perp$ are the vector components of the effective magnetic fields in $\hat{\varphi}$ and $\hat{\theta}$ direction. \textbf{g, h,} Examples of magnetic-field angle $\varphi$ dependent $V_{\text{XY}}^{2\omega}$ (arbitrary units) signals with and without $\tau_{\text{DL}}^{\text{Z}}$. 
        }
    	\label{fig:3}
    \end{figure}

$\hat{\sigma}$ denotes the polarization direction of the spin current.  For high-symmetry spin-source materials, $\hat{\sigma}$ is constrained to lie in the sample plane perpendicular to the charge current, i.e., parallel to our Y axis. For the measurement techniques described in this review we will limit ourselves to cases where the magnetization has only small deflections from the sample plane, in which case for high-symmetry spin-source materials $\vec{\tau}_\text{DL}$ lies in the sample plane and $\vec{\tau}_\text{FL}$ is perpendicular to the plane. For low-symmetry spin-source materials $\vec{\tau}_\text{DL}$ and $\vec{\tau}_\text{FL}$ can each have components both in and out of the sample plane. As long as the polarization direction of the spin current is independent of the magnetization direction and the magnetization undergoes only small deflections from the sample plane, the torque amplitudes ($\tau_\parallel$ in the $\hat{\varphi}$ direction and $\tau_{\perp}$ in the $\hat{\theta}$ direction) are well-described by the angular dependences:
\begin{equation}
\tau_{\parallel}(\varphi)=-\tau^0_{\text{DL,Y}}\cos\varphi-\tau^0_{\text{DL,X}}\sin\varphi+\gamma B^0_{\text{FL,Z}},
\label{tauparallel}
\end{equation}
\begin{equation}
\tau_{\perp}(\varphi)=-\gamma B^0_{\text{FL,Y}}\cos\varphi-\gamma B^0_{\text{FL,X}}\sin\varphi-\tau^0_{\text{DL,Z}}.
\label{tauperp}
\end{equation}
For the coefficients in these expressions, the subscripts DL and FL denote whether the contributions come from the damping-like or field-like torque, and the subscripts X, Y, Z, denote which vector component of $\hat{\sigma}$ contributes.  We use the sign conventions for the torques and magnetic fields illustrated in Fig.~\ref{fig:3}a: a positive value of $\tau^0_{\text{DL,X}}$ corresponds to a torque in the $+\rm X$ direction and a positive value of $\tau^0_{\text{DL,Z}}$ corresponds to a torque in the $+\rm Z$ direction, but a positive value of $\tau^0_{\text{DL,Y}}$ corresponds to a torque in the $-\rm Y$ direction.  This last choice corresponds to a positive conventional SOT efficiency for Pt in a Pt/magnet bilayer with Pt on the bottom.  Similarly the sign conventions for the field-like terms are that a positive value of $B^0_{\text{FL,X}}$ corresponds to a field in the $+\rm X$ direction and a positive value of $B^0_{\text{FL,Z}}$ corresponds to a field in the $+\rm Z$ direction, but a positive value of $B^0_{\text{FL,Y}}$ corresponds to a field in the $-\rm Y$ direction.  By this convention the current-induced in-plane Oersted field contributing to $B^0_{\text{FL,Y}}$ is positive for a metal/magnet bilayer with the metal layer on the bottom.\\

These same torques can alternatively also be expressed in the form of effective magnetic fields of the form $\vec B_{\rm DL} \propto -(\hat{\sigma}\times\hat{m})$ and $\vec B_{\rm FL} \propto \hat{\sigma}$, for which the total current-induced components in the $\hat{\varphi}$ and $\hat{\theta}$ directions have the angular dependences:
\begin{equation}
    \Delta B_\parallel = -B_{\rm FL,Y}^{\rm 0}\cos{\varphi}-B_{\rm FL,X}^{\rm 0}\sin{\varphi}-\frac{\tau_{\rm DL,Z}^{\rm 0}}{\gamma},
\end{equation}
\begin{equation}
    \Delta B_\perp = \frac{\tau_{\rm DL,Y}^{\rm 0}}{\gamma}\cos{\varphi}+\frac{\tau_{\rm DL,X}^{\rm 0}}{\gamma}\sin{\varphi}-B_{\rm FL,Z}^{\rm 0}.
\end{equation}

The terms $\tau^0_\text{DL,Y}$ and $B^0_\text{FL,Y}$ are the only ones allowed for high-symmetry spin-source materials for which $\hat{\sigma} \parallel \pm\rm Y$, while all the others require some type of symmetry breaking. The term $B^0_\text{FL,Y}$ contains the in-plane Oersted field as well as a potential contribution from a conventional field-like SOT. \\

The primary goal of both ST-FMR and second-harmonic Hall measurements is to quantify the strength of SOTs and their associated dimensionless efficiencies per unit applied current density, $\xi_{\rm DL}^{\rm K}$ and $\xi_{\rm FL}^{\rm K}$:
\begin{equation}
    \xi_{\text{DL}}^{\rm K}=\tau_{\text{DL,K}}^0\frac{eM_\text{s}t_{\text{FM}}}{\mu_\text{B}J_\text{e}},
    \label{DL torque}
\end{equation}
\begin{equation}
    \xi_{\text{FL}}^{\rm K}=\gamma B_{\text{FL,K}}^0\frac{eM_\text{s}t_{\text{FM}}}{\mu_\text{B}J_\text{e}},
    \label{FL torque}
\end{equation}
where K is the direction of the spin-orbit torque, $M_{\rm s}$ is the saturation magnetization, $t_{\rm FM}$ is the thickness of the ferromagnet, and $J_{\rm e}$ is the charge current density in the spin source layer.

\bigskip
\noindent \textbf{\Romannum{3}. Requirements for a useful technology}\\

\begin{figure}[t!]
    	\centering
    	\includegraphics[width=0.75\textwidth]{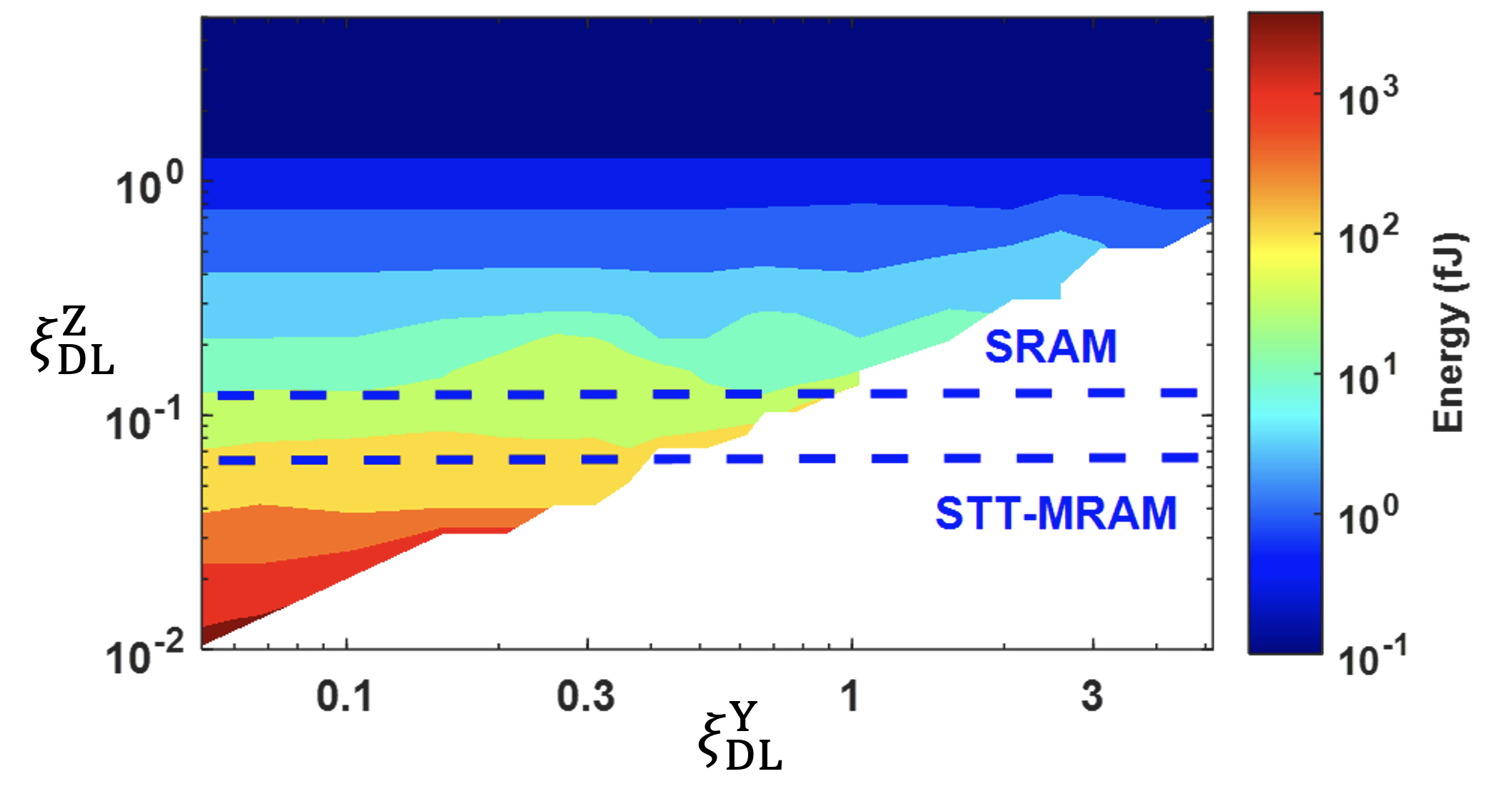}
    	\caption{A calculation of the write energy for a 3-terminal SOT-MRAM device with a 20 nm tunnel junction as a function of out-of-plane and in-plane damping-like torque efficiencies. Adapted from \cite{shazon_TED}.
        }. 
    	\label{fig:2}
\end{figure}

To achieve energy efficient anti-damping switching of a nanomagnet with PMA in an SOT device, both out-of-plane ($\xi_{\text{DL}}^\text{Z}$) and in-plane ($\xi_{\text{DL}}^\text{Y}$) components of the damping-like torque efficiency are required\cite{liao2020spin,leelee2020}.  $\xi_{\text{DL}}^\text{Z}$ determines the current density required for switching by providing the compensation of the intrinsic Gilbert damping, while $\xi_{\text{DL}}^\text{Y}$ reduces the switching time by providing an initial push of the magnetization away from the Z axis where the purely-Z component of the damping-like torque goes to zero. Fig.~\ref{fig:2} shows the calculated write energy for a three-terminal (3T) 20 nm CoFe PMA device\cite{shazon_TED}. As a ball-park number, the write energy of 3T SOT-MRAM should become competitive compared to the existing state-of-the-art magnetic memory technology, spin-transfer-torque MRAM (STT-MRAM)\cite{edelstein2020}  when $\xi_{\text{DL}}^\text{Z}>0.06$\cite{shazon_TED}. A two-terminal (2T) geometry in which out-of-plane damping-like torque from both SOT and conventional STT act in concert can be even more favorable. In an optimized 2T geometry for a 20 nm tunnel junction, modeling suggests that $\xi_{\text{DL}}^\text{Z}>0.051$ can provide a write energy comparable even to static random-access memory (SRAM)\cite{shazon_TED}. To ensure reliable deterministic anti-damping switching, the ratio $\xi_{\text{DL}}^\text{Z}/\xi_{\text{DL}}^\text{Y}$ also needs to be greater than 0.16 for 3T SOT-MRAM, in order to avoid a dynamical regime in which the magnetization is forced to lie along the Y axis rather undergoing switching in the Z direction\cite{liao2020spin}.  This ratio constraint is reduced but still nonzero for the 2T geometries\cite{shazon_TED}. The results of a detailed case study benchmarking the performance of various SOT-MRAM device designs \cite{shazon_TED} is presented in Box 1. \\

\begin{tcolorbox}[
    title=Box 1 | Benchmarking of SOT-MRAM Device Designs,
    colback=gray!5,
    colframe=black,
    boxrule=0.5pt,
    sharp corners]
Here we compare various SOT-MRAM device configurations and benchmark their performance in terms of write energy \cite{shazon_TED}. A 3-terminal SOT-MRAM device consists of a magnetic tunnel junction (MTJ) and an SOT layer that provides both the in-plane ($\tau_{\rm DL}^{\rm Y}$) and out-of-plane ($\tau_{\rm DL}^{\rm Z}$) component of the damping-like SOT (panel \textbf{a}). In a 2-terminal SOT-MRAM device, the write and read currents share the same path and the out-of-plane damping-like torque from both SOT and STT act in concert to switch the free-layer magnetization (panel \textbf{b}). An optimized 2-terminal SOT-MRAM is shown in panel \textbf{c}, where the dimension of the fixed layer is reduced by half. The dimension reduction of the fixed layer increases the SOT efficiency. \\

\includegraphics[width=1\textwidth]{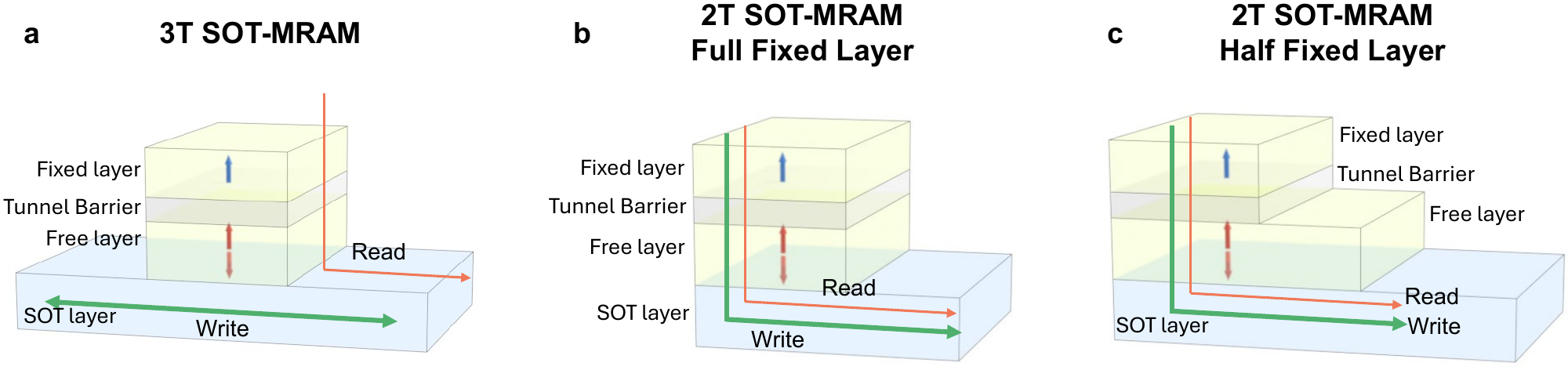}

Table~\ref{tab:sot_mram_requirements} summarizes the  
torque efficiency ($\xi_{\rm DL}^{\rm Z}$) required in order that these three SOT-MRAM device designs become competitive in write energy compared to existing STT-MRAM and SRAM  technologies. 

\captionof{table}{Requirements for SOT-MRAM devices}
\label{tab:sot_mram_requirements}

\vspace{0.3em}

\definecolor{headerbluegray}{RGB}{190,198,211}
\definecolor{rowbluegray}{RGB}{222,227,236}
\definecolor{rowlightbluegray}{RGB}{240,243,248}

\newcommand{\tallcell}[1]{\rule[-1.0em]{0pt}{3.0em}#1}

\renewcommand{\arraystretch}{1.25}
\renewcommand\cellalign{cc}
\renewcommand\theadalign{cc}
\setlength{\tabcolsep}{6pt}

\begin{tabularx}{\textwidth}{%
    >{\centering\arraybackslash}p{0.2\textwidth}
    >{\centering\arraybackslash}X
    >{\centering\arraybackslash}X
    >{\centering\arraybackslash}X
}
\hline

\rowcolor{headerbluegray}
\tallcell{\textbf{Requirements}}
& \tallcell{\textbf{3T SOT-MRAM}}
& \tallcell{\textbf{\makecell{2T SOT-MRAM\\[-0.1em]Full Fixed Layer}}}
& \tallcell{\textbf{\makecell{2T SOT-MRAM\\[-0.1em]Half Fixed Layer}}} \\
\hline

\rowcolor{rowlightbluegray}
\tallcell{Symmetry}
& \multicolumn{3}{c}{\tallcell{$\tau_{\rm DL}^{\rm Y}+\tau_{\rm DL}^{\rm Z}$}} \\
\hline

\rowcolor{rowbluegray}
\tallcell{Mechanism}
& \multicolumn{3}{c}{\tallcell{Anti-damping switching ($\tau_{\rm DL}^{\rm Z}\propto \alpha \gamma B_{\rm anis}$})} \\
\hline

\rowcolor{rowlightbluegray}
\tallcell{\makecell{\small $\xi_{\rm DL}^{\rm Z}$ to surpass\\ STT-MRAM}}
& \tallcell{$0.06$}
& \tallcell{$>0$}
& \tallcell{$>0$} \\
\hline

\rowcolor{rowbluegray}
\tallcell{\makecell{\small $\xi_{\rm DL}^{\rm Z}$ to surpass\\ SRAM}}
& \tallcell{$0.13$}
& \tallcell{$0.1$}
& \tallcell{$0.051$} \\
\hline

\end{tabularx}

\begin{minipage}{0.95\textwidth}
\footnotesize
\textit{Note:} The table is based on MTJ size = 20 nm. $\alpha$ is the effective Gilbert damping constant. $\gamma$ is the gyromagnetic ratio. A conventional damping-like torque efficiency ($\xi_{\rm DL}^{\rm Y}$) of 0.3 is assumed for the simulation. 
\end{minipage}
\end{tcolorbox}

\noindent \textbf{\Romannum{4}. Quantitative characterization methods}\\

\noindent\textbf{A. Spin-torque ferromagnetic resonance (ST-FMR)}\\

\noindent\textbf{a. Explanation of the technique}\\

The first measurement approach we will consider is spin-torque ferromagnetic resonance, or ST-FMR.  To perform ST-FMR, one applies a microwave current with fixed frequency $I=I_{\rm rf}\cos{(\omega t)}$ in the sample plane and measures the d.c.\ voltage as an applied magnetic field $B$ is swept through the magnetic resonance condition. For simplicity, we generally apply the magnetic field within the sample plane, and the microwave frequency should be chosen to be high enough so that near the resonance field the magnetic layer is saturated to be spatially uniform and parallel to the applied field in the absence of charge current. The longitudinal resistance in such a system can then be written as $R_{\rm XX}=R_0+R_{\rm AMR}m_{\rm X}^{2}=R_0+R_{\rm AMR}\cos^2{\varphi}$. The current-induced torques acting on the free-layer ferromagnet excite its magnetic moment away from equilibrium, causing (in the linear regime) the anisotropic magnetoresistance of the magnet to oscillate at the same frequency as the microwave current, $R_{\rm XX}(t)=R_0- R_{\rm AMR}\sin{2\varphi}Re[\delta\varphi e^{-i\omega t}]$ (Fig.~\ref{fig:3}b), where the precession amplitude $\delta\varphi$ near resonance is related to the torques as\cite{karimeddiny2020transverse}:
\begin{equation}
    \delta\varphi=\frac{-\omega_2\tau_\perp+i\omega\tau_\parallel}{-\gamma(B-B_0)\omega^++i\alpha\omega\omega^+}.
\end{equation}
Here $B_0$ is the resonant field, $\omega^+=\omega_1+\omega_2$ with $\omega_1=\gamma B_0$ and $\omega_2=\gamma(B_0+ \mu_0M_{\text{eff}})$, $\mu_0 M_{\text{eff}}$ (T) is the anisotropy field for the thin-film magnet (we take $\mu_0 M_\text{eff}>0$ for samples with in-plane magnetic anisotropy and $\mu_0 M_\text{eff}<0$ for samples with PMA), and $\omega = \sqrt{\omega_1 \omega_2}$ is the resonance frequency.\\

The mixing of the time-dependent resistance with the alternating current produces a d.c.\ voltage, $V_{\text{mix}}=-\frac{I_{\rm rf}}{2}R_{\rm AMR}\sin{2\varphi}Re[\delta\varphi]$, with a resonance shape that is the sum of a symmetric Lorentzian $S(B) = \Delta^2/[\Delta^2 + (B-B_0)^2]$ and an antisymmetric Lorentzian $A(B) = (B-B_0)\Delta/[\Delta^2 + (B-B_0)^2]$ \cite{liu2011spin,macneill2017control,karimeddiny2020transverse,bose2022tilted}:
\begin{equation}
    \label{Vmix}
     V_{\text{XX}}^{\text{mix}} = V_\text{S}S(B) + V_\text{A}A(B). 
\end{equation}
$\Delta = \alpha \omega / \gamma$ is the half-width-at-half-maximum resonance linewidth.
Using Eqs.~(\ref{tauparallel}, \ref{tauperp}), the symmetric and anti-symmetric components of $V_{\text{XX}}^{\text{mix}}$ contributed by the X, Y, and Z components of the torques depend on the magnetic-field angle as:
     \begin{align}
        V_\text{S} &= \frac{I_{\text{rf}}R_{\text{AMR}}}{2\alpha\omega^+}\Big( \tau_{\text{DL,X}}^0 \sin2\varphi \sin\varphi  + \tau_{\text{DL,Y}}^0 \sin2\varphi \cos\varphi - \gamma B_{\text{FL,Z}}^0 \sin2\varphi \Big), \nonumber\\
        &\equiv S^\text{X}_\text{DL}\sin2\varphi \sin\varphi + S^\text{Y}_\text{DL}\sin2\varphi \cos\varphi + S^\text{Z}_\text{FL} \sin2\varphi,
        \label{V_S}
    \end{align}
    \begin{align}
        V_\text{A} &= \frac{I_{\text{rf}}R_{\text{AMR}}}{2\alpha\omega^+}\frac{\omega_2}{\omega} \Big( \gamma B_{\text{FL,X}}^0 \sin2\varphi \sin\varphi + \gamma B_{\text{FL,Y}}^0 \sin2\varphi \cos\varphi + \tau_{\text{DL,Z}}^0 \sin2\varphi \Big),\nonumber\\ 
        &\equiv A^\text{X}_\text{FL}\sin2\varphi \sin\varphi + A^\text{Y}_\text{FL}\sin2\varphi \cos\varphi + A^\text{Z}_\text{DL} \sin2\varphi.
        \label{V_A}
    \end{align}

    By fitting the measured angular dependence of $V_\text{S}$ and $V_\text{A}$ to Eqs.~(\ref{V_S}, \ref{V_A}) the different torque components can be extracted. The relationship between the measured mixing voltages and the strengths of respective torques are summarized in Eq.~(\ref{S_X}-\ref{A_Z}) in Box 2. Example mixing voltages contributed by different torque components as a function of external magnetic-field angle are shown in Fig.~\ref{fig:3}c, d.\\
    
    The most challenging step in making quantitative ST-FMR measurements of spin-torque efficiencies is often the determination of the microwave current amplitude $I_\text{rf}$ within the sample. In our group's first use of ST-FMR\cite{liu2011spin}, we attempted to perform this calibration by assuming that the field-like torque was due entirely to the in-plane Oersted field, so that the torque efficiency could be determined directly from the ratio $S^\text{Y}_\text{DL}/A^\text{Y}_\text{FL}$ (i.e., the ``S/A'' ratio). However, this is generally incorrect, because field-like SOTs can sometimes be comparable to or larger than the Oersted torque\cite{fan2013,skinner2014,pai2015dependence}. For quantitative measurements of torque efficiencies, one should be more careful. We have had success with two methods: direct calibrations of $I_\text{rf}$, and studies using samples with different layer thicknesses to separate interfacial field-like SOTs from volume-acting Oersted torques.  The details of these procedures are discussed in the Supplementary Information.\\
    
    An important general check to help verify the reality of unconventional SOT components is to perform magnetic-field angle dependent ST-FMR measurements on devices oriented along various crystallographic axes to establish the dependence of the torque symmetry on the crystallographic axes as well as the magnetic-field angle.  The dependences for the amplitudes of the different torque components as a function of the angle of the applied electric field relative to crystal axes are generally constrained by the sample symmetries \cite{macneill2017control,bose2022tilted,patton2023symmetry,huang2025control}.\\

\begin{tcolorbox}[
    title=Box 2 | Experimental Observables in ST-FMR,
    breakable,
    colback=gray!5,
    colframe=black,
    boxrule=0.5pt,
    sharp corners]
\setcounter{boxeq}{0}

For the ST-FMR technique, the measured fitting coefficients $S^\text{X}_\text{DL}$, $S^\text{Y}_\text{DL}$, and $S^\text{Z}_\text{FL}$ (see Eq.~(\ref{V_S})) and $A^\text{X}_\text{FL}$, $A^\text{Y}_\text{FL}$, and $A^\text{Z}_\text{DL}$ (see Eq.~(\ref{V_A})) are related to the X, Y, and Z component of the damping-like and field-like torques according to:
    \begin{equation}
     S_{\text{DL}}^\text{X}=\frac{I_{\text{rf}}R_{\text{AMR}}}{2\alpha\omega^+}\tau_{\text{DL,X}}^0,
     \boxeqtag
     \label{S_X}
    \end{equation}
    \begin{equation}
    S_{\text{DL}}^\text{Y}=\frac{I_{\text{rf}}R_{\text{AMR}}}{2\alpha\omega^+}\tau_{\text{DL,Y}}^0, 
    \boxeqtag
     \label{S_Y}
    \end{equation}
    \begin{equation}
     S_{\text{FL}}^\text{Z}=-\frac{I_{\text{rf}}R_{\text{AMR}}}{2\alpha\omega^+}\gamma B_{\text{FL,Z}}^0,
     \boxeqtag
     \label{S_Z}
    \end{equation}
    \begin{equation}
    A_{\text{FL}}^\text{X}=\frac{I_{\text{rf}}R_{\text{AMR}}}{2\alpha\omega^+}\frac{\omega_2}{\omega}\gamma B_{\text{FL,X}}^0,
    \boxeqtag
     \label{A_X}
    \end{equation}
    \begin{equation}
     A_{\text{FL}}^\text{Y}=\frac{I_{\text{rf}}R_{\text{AMR}}}{2\alpha\omega^+}\frac{\omega_2}{\omega}\gamma B_{\text{FL,Y}}^0, 
     \boxeqtag
     \label{A_Y}
    \end{equation}
    \begin{equation}
    A_{\text{DL}}^\text{Z}=\frac{I_{\text{rf}}R_{\text{AMR}}}{2\alpha\omega^+}\frac{\omega_2}{\omega}\tau_{\text{DL,Z}}^0. \\ 
    \boxeqtag
     \label{A_Z} \\
    \end{equation} 
The negative sign in Eq.~(\ref{S_Z}) represents a correction to ref.~\cite{karimeddiny2020transverse}.

\end{tcolorbox}

\noindent \textbf{b. Common artifacts and their exclusion from ST-FMR measurements}\\

There are several artifacts which can interfere with ST-FMR measurements, requiring care in the analysis.  \\

\romannum{1}. \textbf{Spin Pumping}\\

Spin pumping together with the inverse spin Hall effect (ISHE)\cite{tserkovnyak2002enhanced,tserkovnyak2002spin,mosendz2010detection} and resonant heating together with the longitudinal spin Seebeck effect and the ISHE \cite{holanda2017longitudinal} can produce symmetric resonance peak shapes which have the same dependence on $\varphi$ as the mixing voltage from the conventional in-plane damping-like SOT ($\xi^\text{Y}_\text{DL}$). These effects can therefore interfere with measurements of the conventional damping-like SOT efficiency, but they are easily distinguished from the measurements of $\xi^\text{Z}_\text{DL}$ by a different dependence on the magnetic-field angle. The spin pumping signal becomes increasingly important (relative to the SOT signal) with thicker sample layers -- either or both a thick spin-source layer or a thick detecting ferromagnet. As a rough rule of thumb, the effects of spin pumping should not be ignored for samples with electrically-conducting ferromagnets in which the thickness of either the spin-source layer or the detecting ferromagnet approaches 8-10 nm\cite{karimeddiny2020transverse,cham2022separation}, with even more stringent conditions for electrically-insulating ferromagnets. For situations in which spin pumping or resonant heating are significant, the conventional spin-torque signal can still be separated from the artifact signals by measuring mixing voltages both longitudinally and transversely\cite{karimeddiny2020transverse}.\\

\romannum{2}. \textbf{Device Asymmetry} \\

Our standard measurement geometry employs a ground-signal-ground probe to make contact to a device structure with bilateral symmetry (See Fig.~\ref{fig:3}a) so that by symmetry the net out-of-plane Oersted field in the device region produced by current flow feeding into and out of the device is zero. A non-symmetric current flow can produce a non-zero net out-of-plane Oersted field at the position of the device. This corresponds to an in-plane field-like torque that contributes to the symmetric mixing voltage with a $\sin(2\varphi)$ symmetry. Device asymmetry can be caused by flaws in device design and fabrication including non-uniform etching, photoresist residue, or a dirty or bent ground-contact probe. This type of artifact can be confused with an unconventional Z-type field-like torque ($\xi^\text{Z}_\text{FL}$), but it usually lacks consistency across different devices.\\

\romannum{3}. \textbf{Substrate Conduction}\\

When substrates have some non-zero conductivity at microwave frequencies, the unwanted microwave current generated in the substrate is often not uniform, so that it can produce a net out-of-plane Oersted field as well as an in-plane Oersted field. The extra in-plane Oersted field can add to the Oersted field produced by the spin-source material, thereby altering measurements of the conventional field-like torque. The extra out-of-plane Oersted field produces an in-plane field-like torque that mimics the symmetry of a $\xi^\text{Z}_\text{FL}$ component. One common substrate that is often problematic is SrTiO$_3$, which becomes conductive to varying degrees if ion milling is used during device fabrication, due to the formation of oxygen vacancies\cite{reagor2005highly}. Therefore, when fabricating devices on SrTiO$_3$ substrates for SOT measurements, exposure of the SrTiO$_3$ surface to ion milling  should be avoided. If ion milling is unavoidable, chemical etching should be used afterward to remove the last layer at the top surface of the substrate. For example, NaIO$_4$ can effectively oxidize ruthenium in SrRuO$_3$ from Ru$^{4+}$ to water soluble Ru$^{8+}$\cite{pan2023defect}. When chemical etching is not effective, post-annealing of the substrate in an oxygen environment can also help to minimize the harmful effects of oxygen vacancies. One caveat for post-annealing is that this could potentially induce unwanted oxidation and inter-diffusion in the whole material stack.\\

\romannum{4}. \textbf{Current-induced magnon creation and annihilation}\\

As we will discuss in greater detail below, No\"el \textit{et al.}\cite{noel2025nonlinear,noel2025estimation} have shown that current-induced magnon creation and annihilation can cause a failure of the assumption that the magnitude of the magnetization remains constant independent of the applied current, and the resulting magnon magnetoresistance can produce substantial artifacts affecting second-harmonic Hall measurements.
The standard ST-FMR analysis also assumes that the magnitude of the magnetization vector is independent of the applied current, so it is important to consider whether this magnon magnetoresistance might also affect ST-FMR measurements. We assert that the non-resonant magnon processes considered by refs.~\cite{noel2025nonlinear,noel2025estimation} should not significantly change ST-FMR results -- non-resonant processes in general might contribute to a smoothly-varying background in ST-FMR, but they should not directly modify the resonant voltage peak that is analyzed to determine torque efficiencies. However, a population of excited short-wavelength magnons might alternatively be created resonantly, by scattering of magnons from the uniform mode that is driven during ST-FMR to modes with non-zero wavenumber. This might alter the value of $M_\text{s}$ to be used in the ST-FMR analysis slightly relative to the equilibrium value. Nevertheless, we would still not expect it to produce a separate artifact signal because standard ST-FMR measurements do not include a substantial applied direct current that would transduce an associated change in magnon magnetoresistance to a voltage signal.  \\

However, a large direct current is applied during the technique of dc-biased ST-FMR, in which damping-like torques are measured from the change in resonant linewidth as a function of d.c.\ current\cite{ando2008,karimeddiny2021}, so that in this technique a current-dependent magnon magnetoresistance could alter the results. We speculate that, because of a magnon magnetoresistance, a resonant excitation of magnons might pollute dc-biased ST-FMR measurements by creating a voltage peak or dip with the form $\propto I_\text{dc} \Delta M(B)$, where $\Delta M(B)$ has a resonance-like lineshape with a current-dependent linewidth that differs from the primary ST-FMR resonance\cite{karimeddiny2021}. Our experience is that the results of dc-biased ST-FMR sometimes do differ by large amounts from conventional ST-FMR for some types of samples, and so we consider dc-biased ST-FMR to be a less reliable technique.  A clever variant of ST-FMR introduced by Kobayashi \textit{et al.}\cite{kobayashi2024}, might also potentially be affected by this type of resonantly-excited magnon magnetoresistance.\\

\noindent\textbf{B. Second-harmonic Hall measurements}\\

\noindent\textbf{a. Explanation of the technique}\\

The second technique commonly used to quantify current-induced SOTs is second-harmonic Hall voltage measurements, in which a low-frequency (i.e., non-resonant) alternating current $I=I_0\sin{\omega t}$ is applied at a frequency $\omega$ 
and a transverse Hall voltage is detected. 
The Hall resistance of the system can be written as: $R_{\rm XY}=R_{\rm AHE}m_{\rm Z}+R_{\rm PHE}m_{\rm X}m_{\rm Y}=R_{\rm AHE}\cos{\theta}+\frac{1}{2}R_{\rm PHE}\sin^2 \theta \sin{2\varphi}$.\footnote{The literature is not always consistent about factors of 2 for this convention. Some papers write the anomalous Hall contribution as $\frac{1}{2}R_{\rm AHE}^*\cos{\theta}$ and others write the planar Hall contribution as $R_{\rm PHE}^*\sin^2 \theta\sin{2\varphi}$.} The applied alternating current drives the magnetization of the detecting ferromagnet away from its equilibrium position, producing planar and anomalous Hall resistances that oscillate at  $\omega$ ($\varphi=\varphi+\delta\varphi\sin{\omega t}$, $\theta=\theta+\delta\theta\sin{\omega t}$).  Unlike in ST-FMR measurements, the frequency of the drive current in second-harmonic Hall measurements is generally sufficiently small that the deflections of the magnetization are quasistatic, so that the oscillations of $\varphi$ and $\theta$ are in-phase with the drive current.  The mixing of the time-dependent Hall resistances with the applied alternating current yields a Hall voltage of the form $V_{\rm XY}=R_{\rm XY}(t)I(t)=I_0R_{\rm XY}^0\sin{\omega t}+I_0\frac{dR_{\rm XY}}{d\varphi}\delta\varphi\sin^2{\omega t}+I_0\frac{dR_{\rm XY}}{d\theta}\delta\theta\sin^2{\omega t}$, producing mixing-voltage components at both d.c.\ and $2\omega$. For an in-plane magnetized system, $\delta\varphi$ and $\delta\theta$ can be related to the current-induced effective magnetic fields as\cite{hayashi2014quantitative}:
\begin{equation}
    \delta\varphi=\frac{\Delta B_\parallel}{B-B_{\rm A}\cos{2\varphi}},
\end{equation}
\begin{equation}
    \delta\theta=\frac{\Delta B_\perp}{B+\mu_0M_{\rm eff}+B_{\rm A}\sin^2{\varphi}},
\end{equation}
where the term $B_{\rm A}$ allows for an in-plane anisotropy field favoring the $\rm Y$ direction.\\

Experiments focus on the $2\omega$ Hall component in an attempt to limit artifacts. A typical device geometry for second-harmonic Hall measurements is shown in Fig.~\ref{fig:3}e. Assuming that the in-plane anisotropy $\propto B_{\rm A}$ is sufficiently weak relative to $B$ to be neglected, the second-harmonic Hall voltage $V_{\text{XY}}^{2\omega}$ can be expressed as\cite{pi2010tilting,garello2013symmetry,hayashi2014quantitative,avci2014interplay}:
\begin{equation}
\label{eq:3}
\begin{aligned}
V_{\text{XY}}^{2\omega} &=-\frac{1}{2}I_0R_{\rm PHE}\frac{\Delta B_\parallel}{B}\cos{2\varphi}+\frac{1}{2}I_0R_{\rm AHE}\frac{\Delta B_\perp}{B+\mu_0 M_\text{eff}}\\
&= D_{\text{X}} \sin \varphi + D_{\text{Y}} \cos \varphi + D_{\text{Z}} \cos 2\varphi \\
&\quad + F_{\text{X}} \sin \varphi \cos 2\varphi + F_{\text{Y}} \cos \varphi \cos 2\varphi + F_{\text{Z}} + T_{\rm IP} \sin 2\varphi,
\end{aligned}
\end{equation}
where $D_{\text{X}}$, $D_{\text{Y}}$, and $D_{\text{Z}}$ are the contributions to $V_{\text{XY}}^{2\omega}$ produced by the X, Y, Z components of the damping-like torque; $F_{\text{X}}$, $F_{\text{Y}}$, and $F_{\text{Z}}$ are the contributions from field-like components (see Box 3).  The term $T_{\rm IP}\sin 2\varphi$ accounts for a planar Nernst background voltage that can be induced by an in-plane longitudinal thermal gradient\cite{avery2012observation, avery2012determining,karimeddiny2020transverse}. Example second-harmonic Hall voltages with only conventional Y-component contributions ($\xi^\text{Y}_\text{DL}$ and $\xi^\text{Y}_\text{FL}$) and with both Y and Z damping-like components are shown in Fig.~\ref{fig:3}g and Fig.~\ref{fig:3}h respectively.\\
\bigskip
\begin{tcolorbox}[
    title=Box 3 | Experimental Observables in Second-Harmonic Hall Measurements,
    breakable,
    colback=gray!5,
    colframe=black,
    boxrule=0.5pt,
    sharp corners]
\setcounter{boxeq}{0}

For the second-harmonic Hall technique, the measured fitting coefficients $D_{\rm X}$, $D_{\rm Y}$, $D_{\rm Z}$, $F_{\rm X}$, $F_{\rm Y}$, $F_{\rm Z}$ (see Eq.~(\ref{eq:3})) are related to the the X, Y, and Z components of the damping-like and field-like torques according to: 
\begin{equation}
    D_{\text{X}} = \frac{\tau_{\text{DL,X}}^{\text{0}}}{\gamma}\frac{I_0R_{\rm AHE}}{2(B+\mu_0 M_\text{eff})},
    \boxeqtag
\label{Dx}
\end{equation}
\begin{equation}
    D_{\text{Y}} = \frac{\tau_{\text{DL,Y}}^{\text{0}}}{\gamma}\frac{I_0R_{\rm AHE}}{2(B+\mu_0 M_\text{eff})}+V_{\text{ANE}}+V_{\text{ONE}}B,
    \boxeqtag
\label{Dy}
\end{equation}
\begin{equation}
    D_{\text{Z}} = \frac{\tau_{\text{DL,Z}}^{\text{0}}}{\gamma}\frac{I_0R_{\rm PHE}}{2B},
    \boxeqtag
\label{Dz}
\end{equation}
\begin{equation}
    F_{\rm X} = B^0_{\text{FL,X}}\frac{I_0R_{\rm PHE}}{2B},
    \boxeqtag
\label{Fx}
\end{equation}
\begin{equation}
    F_{\text{Y}} = B^0_{\text{FL,Y}}\frac{I_0R_{\rm PHE}}{2B},
    \boxeqtag
\label{Fy}
\end{equation}
\begin{equation}
    F_{\text{Z}} = -B^0_{\text{FL,Z}}\frac{I_0R_{\rm AHE}}{2(B+\mu_0 M_\text{eff})} + C.
    \boxeqtag
\label{Fz}
\end{equation}
Here $R_{\rm AHE}$ is the anomalous Hall resistance ($R_{\rm XY}=R_{\rm AHE}\cos(\theta)$), $R_{\rm PHE}$ is the planar Hall resistance ($R_{\rm XY} = R_{\rm PHE}\sin{\varphi}\cos{\varphi}$), and $I_0$ is the peak amplitude of the applied alternating current. $V_{\text{ANE}}$ is the anomalous Nernst voltage, $V_{\text{ONE}}B$ is the ordinary Nernst voltage, and C accounts for a constant background. The signs are corrected relative to ref.~\cite{bose2022tilted}.
\end{tcolorbox}

To determine the SOT vector components $\tau^0_\text{DL,K}$ and $B^0_\text{FL,K}$, the Hall resistances ($R_\text{PHE}$ and $R_\text{AHE}$) are calibrated, the second-harmonic Hall signal ($V^{2\omega}_\text{XY}$) is measured versus $\varphi$  for a sequence of magnitudes for the applied in-plane magnetic field, then the coefficients $D_\text{X}$, $D_\text{Y}$, $D_\text{Z}$, $F_\text{X}$, $F_\text{Y}$, and $F_\text{Z}$ are determined from fits to the angular dependence in Eq.~(\ref{eq:3}), and finally appropriate fits versus $B$ in Eqs.~(\ref{Dx})-(\ref{Fz}) in Box 3 are performed for each component.\\

\noindent \textbf{b. Common artifacts and their exclusion from second-harmonic Hall measurements}\\

Like ST-FMR measurements, the second-harmonic Hall technique can also be affected by device asymmetry and substrate conduction (although substrate conduction in oxide substrates at low frequencies is typically not as problematic as at microwave frequencies). In addition, second-harmonic Hall voltage measurements are subject to other artifacts that, if not carefully excluded, may compromise the validity of the SOT analysis.\\

\romannum{1}. \textbf{Current-induced magnon creation and annihilation}\\

When using second-harmonic Hall voltage to estimate the strength of SOT, a standard assumption has been that the magnetization of the detecting ferromagnet remains unchanged upon the applying of large ac currents. This is not necessarily true as the creation and annihilation of magnons induced by spin currents can affect the magnitude of the magnetization of the detecting ferromagnet\cite{demidov2011control,demidov2017magnetization,borisenko2018relation,thiery2018nonlinear}, especially for magnets with low Gilbert damping, small saturation magnetization, and low Curie temperature. Magnon creation and annihilation generated by damping-like SOT from a non-resonant charge current acting on an in-plane magnet should cause the magnetization magnitude to have a dependence on $\varphi$ of the form\cite{noel2025nonlinear,noel2025estimation}: $M(I) = M_\text{s} + \Delta M(I)\sin\varphi$. As a consequence, all magnetoresistances that depend on the magnitude of the magnetization, including the planar Hall effect and the anomalous Hall effect, will depend on the current density, and therefore in the presence of an applied current they will produce artifact second-harmonic voltage signals\cite{noel2025nonlinear,noel2025estimation}. For the magnetoresistances pertinent to the interpretation of the second-harmonic Hall measurements, $R_\text{PHE} \propto M_\text{X}M_\text{Y}$ and $R_\text{AHE}\propto M_\text{Z}$. \\

To first order in $\Delta M$, the additional terms in the second-harmonic Hall signal due to magnon creation and annihilation have the angular dependences\cite{noel2025nonlinear,noel2025estimation}: 
\begin{align}
    \Delta V_\text{XY}^{2\omega} &=
     2V_\text{PHE}^{2\omega}(B)\sin^2(\varphi)\cos(\varphi)+V_\text{AHE}^{2\omega}(B)\sin(\varphi)\cos(\theta) \\
    &=V_\text{PHE}^{2\omega}(B)\cos(\varphi)-V_\text{PHE}^{2\omega}(B)\cos(2\varphi)\cos(\varphi)+V_\text{AHE}^{2\omega}(B)\sin(\varphi)\cos(\theta)
    \label{V_2PHE}
\end{align}
with 
\begin{equation}
    V_\text{PHE}^{2\omega}(B)=-\frac{1}{2}\frac{R_\text{PHE}}{M_\text{s}}\frac{dM}{dI} I_\text{ac}^2,
\end{equation}
\begin{equation}
    V_\text{AHE}^{2\omega}(B)=-\frac{1}{2}\frac{R_\text{AHE}}{M_\text{s}}\frac{dM}{dI} I_\text{ac}^2,
\end{equation}
and $I_\text{ac}$ the amplitude of the applied low-frequency alternating current.\\

For a magnetic-field swept in-plane, $\cos \theta \approx 0$, so the changing anomalous Hall resistance should not affect the measurements significantly, but the changing planar Hall resistance should be included in the expressions for $D_\text{Y}$ and $F_\text{Y}$: 
\begin{equation}
    D^\text{revised}_{\text{Y}} = \frac{\tau_{\text{DL,Y}}^{\text{0}}}{\gamma}\frac{V_{\text{AHE}}}{2(B+\mu_0 M_\text{eff})}+V_{\text{ANE}}+V_{\text{ONE}}B+V_\text{PHE}^{2\omega}(B),
\label{Dy_corrected}
\end{equation}
\begin{equation}
    F^\text{revised}_{\text{Y}} = B^0_{\text{FL,Y}}\frac{V_{\text{PHE}}}{2B}-V_\text{PHE}^{2\omega}(B).
\label{Fy_corrected}
\end{equation}
The contribution $V_\text{PHE}^{2\omega}(B)$ is magnetic-field dependent, and therefore current-induced magnon creation and annihilation can interfere with the fits as a function of $B$ required for an accurate determination of $\tau^0_\text{DL,Y}$ and $B^0_\text{FL,Y}$ and hence the conventional torque efficiencies $\xi^\text{Y}_\text{DL}$ and $\xi^\text{Y}_\text{FL}$. \\

The effects of $V_\text{PHE}^{2\omega}(B)$ are responsible for deviations from expected behavior often seen at relatively low magnetic fields when plotting $D_\text{Y}$ as a function of $(B+\mu_0 M_\text{eff})^{-1}$ and $F_\text{Y}$ as a function of $B^{-1}$, deviations that cannot be explained by a failure to saturate the magnetization or to overcome the magnetic anisotropy of the detecting ferromagnet \cite{avci2014interplay,lau2017spin,neumann2018influence}.
No\"el \textit{et al.}\cite{noel2025estimation} have presented one approach for separating out the 
magnonic contribution from $V_\text{PHE}^{2\omega}$ to $D_\text{Y}$ and $F_\text{Y}$ by measuring second-harmonic voltages both transversely and longitudinally over a range of in-plane applied magnetic field magnitudes. They have shown that neglect of magnon creation and annihilation can cause errors in the determination of $\xi^\text{Y}_\text{DL}$ by a factor of 2 in W/CoFeB samples and by an order of magnitude in YIG (yttrium iron garnet)/Pt samples.
An alternative approach is to perform second-harmonic Hall measurements while rotating the magnetic field out-of-plane (in the X-Z plane) so that there is no significant component of magnetization along the Y axis sensitive to magnon creation or annihilation by the conventional Y component of SOT\cite{xuAlternativeHH}. This should work well for measuring conventional SOTs in samples for which unconventional components are absent, but significant unconventional damping-like toques in either the Z or X direction could still produce magnon magnetoresistances that interfere with accurate torque measurements using this strategy.\\

\romannum{2}. \textbf{Geometry of the Hall bar devices}\\

For second-harmonic Hall measurements, another potential source of inaccuracy is the geometry of the Hall bar devices. The ratio between the width of the Hall probes ($w_\text{H}$) and the main current channel ($w_\text{L}$) determines the uniformity of the current distribution near the Hall cross area. A large ratio causes current leakage into the voltage leads. This does not matter for measuring the linear Hall resistivity from the first-harmonic Hall signal (for a spatially-uniform material), but current non-uniformities will alter the second-harmonic Hall signal, resulting in an underestimate of the torque efficiencies\cite{neumann2018influence}. For accurate torque measurements, $w_\text{H}/w_\text{L} \le 0.2$ is reasonable\cite{neumann2018influence}. \\

\bigskip
\noindent \textbf{\Romannum{5}. Qualitative characterization methods}\\

We also wish to comment briefly about other techniques which have sometimes been applied to the study of unconventional components of SOT.\\

\textbf{A. Field-free magnetization switching}\\

We have discussed that an out-of-plane damping-like torque can efficiently switch the magnetization of a PMA nanomagnet by compensating the Gilbert damping of the system\cite{liao2020spin} until the effective damping is negative, at which point the magnetization of the nanomagnet can spiral away from its initial equilibrium position with increasing precession angle until full reversal is achieved (Fig.~\ref{fig:1}c)\cite{ralph2008spin}. Achieving this efficient anti-damping switching process is our goal in pursuing out-of-plane damping-like torques, but as noted below this requires a reasonably strong out-of-plane damping-like torque, $\xi^0_\text{DL,Z}$ greater than about 0.05 or 0.06 in order to provide better energy efficiency than conventional STT-MRAM.  This type of efficient anti-damping switching using out-of-plane damping-like torque has not yet been demonstrated in any prototype nanoscale SOT device. \\

\begin{figure}[t!]
    	\centering
    	\includegraphics[width=0.8\textwidth]{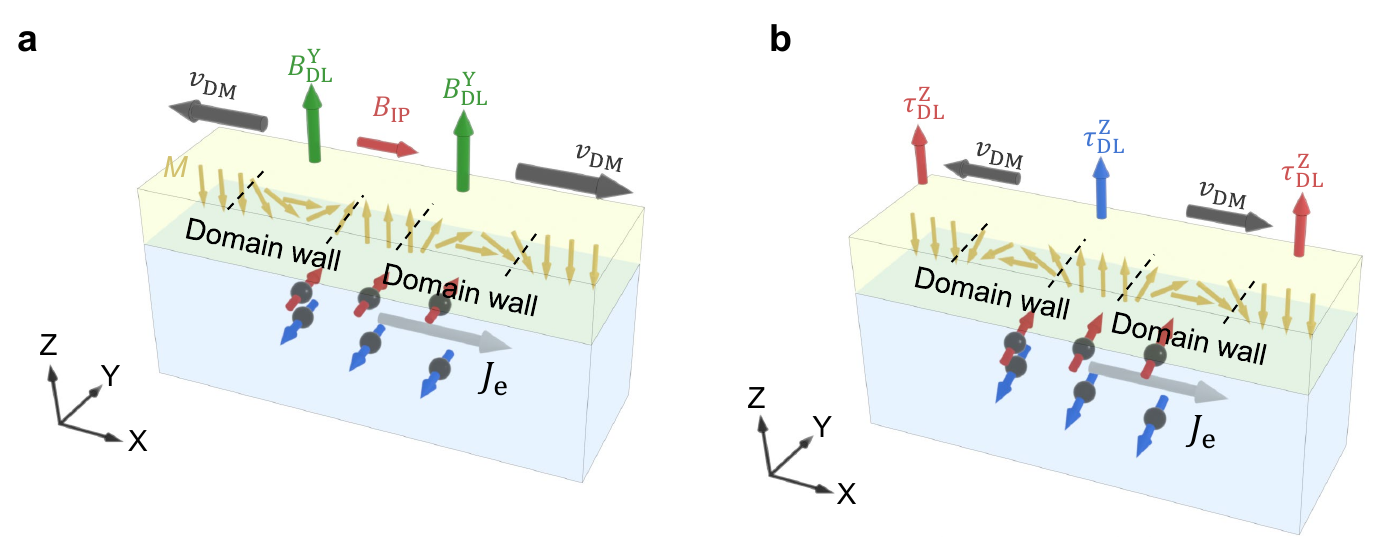}
    	\caption{\textbf{Alternative mechanisms of symmetry breaking for driving switching via domain-wall motion.} \textbf{a,} Symmetry breaking by an external magnetic field applied along the current direction that aligns the magnetic moments inside the domain walls. \textbf{b,} Symmetry breaking by an out-of-plane component of the damping-like torque acting to enhance or suppress thermal fluctuations of the magnetic moments on either side of a domain wall.  Either mechanism allows spin-orbit torque to expand the up-oriented central magnetic domain.  In panel \textbf{b} the arrows in red indicate out-of-plane damping-like torque that enhances thermally-induced magnetization fluctuations in domains with downward-pointing magnetization. The arrow in blue indicates out-of-plane damping-like torque that suppresses thermally-induced magnetization fluctuations in the domain with upward-pointing magnetization.
        }
    	\label{fig:4}
    \end{figure}

In contrast, there is a large literature about using unconventional SOTs to enable ``field-free'' switching in much larger, micron-scale or above, PMA devices (e.g. \cite{liu2021symmetry,chen2021observation,dc2023observation,kao2022deterministic,wang2024field,liu2023field,zhang2023room}). It should be understood that the switching mechanism in that case is not the same as the anti-damping switching of a nanomagnet. In micron-scale PMA samples, magnetization reversal is achieved by domain nucleation and expansion via domain wall motion (Fig.~\ref{fig:4}a)\cite{emori2013current,lee2014central}. This process is driven primarily by the conventional in-plane damping-like SOT, and (unlike anti-damping switching of a nanomagnet) it requires a symmetry-breaking mechanism in order for the switching to be deterministic.  The symmetry breaking can be provided by an externally-applied magnetic-field with a component parallel to the charge-current axis (Fig.~\ref{fig:4}a) or by an unconventional component of SOT (Fig.~\ref{fig:4}b). In either case, for materials studied to date the critical current density appears still to be determined primarily by the strength of the conventional component of the SOT, so that the value of the critical current density does not provide a quantitative measure of the unconventional torque component that merely provides the symmetry breaking (Fig.~\ref{fig:4}b)\footnote{Actually, due to variations in processes of domain nucleation and domain wall pinning, the current densities required for switching in micron-scale PMA devices are not even a good predictor of the conventional in-plane damping-like torque efficiency, see ref.~\cite{zhumicron}.}.  Therefore, an observation of ``field-free'' spin-orbit switching in micron-scale samples by itself should not be used to claim that a low-symmetry spin-source material has promise for generating efficient switching of practical nanoscale magnetic memory devices that are too small to host domain walls.  It is for this reason that we consider accurate quantitative measurements of the out-of-plane damping-like torque to be an essential first step for the identification of spin-source materials which might be suitable for switching nanoscale PMA magnetic devices, prior to the full fabrication of device prototypes. \\

\textbf{B. Out-of-plane magnetic hysteresis loop shift}\\

Pai \textit{et al.}\cite{paibeach2016} introduced a convenient and reliable technique to measure the conventional in-plane damping-like torque in micron-scale (heavy metal)/(PMA magnet) samples based on current-induced shifts in magnetic hysteresis loops measured as a function of sweeping an out-of-plane magnetic field for a sequence of different values of in-plane magnetic-field (in the X direction) (Fig.~\ref{fig:4}a).  In some samples with a reduced symmetry induced by an angled deposition (i.e., ``wedged'' samples), ref.~\cite{paibeach2016} found that the current-induced out-of-plane effective field was non-zero even at zero applied in-plane magnetic field and linear in the applied current. This is an interesting consequence of reduced symmetry, but given the geometry of the experiment the offset does not correspond to the type of out-of-plane damping-like torque we seek for efficient switching of nanomagnets, but rather it corresponds to an out-of-plane field-like term ($\xi^\text{Z}_\text{FL}$) influencing domain-wall motion.  Based on optical imaging of the switching process, Pai \textit{et al.}\ also suggested the possibility that the offset in the lowered-symmetry samples might not correspond to an unconventional current-induced field at all, but rather might be due to an asymmetry in the positions of domain nucleation within the sample\cite{paibeach2016}.\\

Hysteresis loop-shift measurements on micron-scale samples have more recently been connected with out-of-plane damping-like torques based on loop shifts that are non-linear in the applied current, rather than linear -- typically there is negligible loop shift below some threshold current magnitude, at which point the loop shift turns on\cite{baek2018spin}. We agree that the nonlinear-in-current loop shifts are good qualitative evidence for the existence of a non-zero out-of-plane damping-like torque, but nevertheless this approach does not provide a straightforward strategy for making quantitative measurements of this torque component. Our interpretation of these experiments is that in a micron-scale PMA sample containing domain walls the effect of an out-of-plane damping-like torque is to enhance the amplitude of thermally-induced magnetization fluctuations for the out-of-plane magnetization on one side of each domain wall, and to suppress them on the other side (Fig.~\ref{fig:4}b). This tends to stabilize the magnetization orientation on the side with decreased fluctuations and enhance the probability that the domain wall will shift in the direction of the enhanced fluctuations (Fig.~\ref{fig:4}b).  After the domain wall shifts, the out-of-plane damping-like torque will then help stabilize the magnetization on the trailing edge of the shifted domain wall and keep it from shifting back.\\ 

This mechanism provides a symmetry breaking that can appear qualitatively like an out-of-plane magnetic field with a sign that depends on the sign of current (and it can allow for deterministic switching of micron-scale samples in which the domain wall motion is driven primarily by the conventional in-plane SOT), but the consequences on magnetic fluctuations of an energy non-conserving damping-like torque are not quantitatively the same as the application of an energy-conserving magnetic field.  In fact, for a thermally activated process like domain-wall depinning with dynamics proportional to $\exp{(-E/k_{\rm B} \tilde{T})}$, the effects of a damping-like torque are best understood as modifying the effective temperature $\tilde{T}=T/(1-I/I_\text{C})$ (where $I_\text{C}$ is the current where the damping is fully compensated to zero)\cite{koch2004,Li2004activation}. Consequently, the dependence of the loop shift on current $\Delta B_\perp(I)$ is not simply equal to $\tau^0_\text{DL,Z}(I)/ \gamma$, and  it is incorrect to assume that this relationship allows $\xi_{\rm DL}^{\rm Z}$ to be determined from the slope $d(\Delta B_\perp)/dI$. Furthermore, contrary to an initial interpretation\cite{baek2018spin}, we suggest that the threshold current in the hysteresis loop shift experiments does not correspond to the value for which the effective damping is fully compensated so as to change sign from positive to negative, but rather it is merely the value for which the enhanced magnetization fluctuations caused by the out-of-plane damping-like torque begin to dominate over the intrinsic thermal fluctuations in promoting the unpinning of domain walls, at which point the overall effective damping is still positive.\\

\textbf{C. Optical measurements of spin-orbit torques}\\

As explained in our discussion above, both of the electrically-based techniques used for measuring unconventional SOTs (ST-FMR and the second-harmonic Hall technique) are subject to artifacts about which one must be careful when making measurements.  In principle, unconventional torques can also be quantified by using optical methods to measure the amount and direction of current-induced magnetic deflection. Measurements of the conventional  component of in-plane damping-like torque have been demonstrated using magneto-optical Kerr effect (MOKE) imaging\cite{fanMOKE2014,montazeriMOKE,fanMOKE2016,tsai2018,marui2018,chen2021optic,shiota2021} and Sagnac interferometry\cite{Sagnac}. Only a small number of experiments have performed optical measurements of unconventional torques components\cite{humphries2017observation,wangSpinrotation2020,wangPRMaterials2023}.  This is an interesting direction for further development.

\bigskip
\noindent \textbf{\Romannum{6}. Our experimental recommendations}\\

To obtain a reliable measurements of the SOT efficiency (for both the conventional and unconventional torque components), we suggest doing the following.\\

\noindent\textbf{A. For the ST-FMR technique:}\\

\romannum{1}. Use thin detecting ferromagnet and spin-source layers (much thinner than 10 nm) to minimize spin pumping. Alternatively, measure both transverse and longitudinal mixing voltages and correct for these artifact contributions as discussed in ref.~\cite{karimeddiny2020transverse}.\\

\romannum{2}. Be careful to calibrate the rf current, rather than simply using the ``S/A" approach from ref.~\cite{liu2011spin}.  This calibration can be done either by directly measuring $I_\text{rf}$ or by independently determining the strength of the field-like SOT. (See the Supplemental Information.)\\

\romannum{3}. When calculating the out-of-plane damping-like torque $\tau^0_\text{DL,Z}$ from the ST-FMR amplitude $A^\text{Z}_\text{DL}$, recognize that the formula for doing so is different than the formula for calculating the in-plane damping-like torque $\tau^0_\text{DL,Y}$ from $S^\text{Y}_\text{DL}$. (Note the extra factor of $\frac{\omega_2}{\omega}$ in Eq.~(B6) compared to Eq.~(B2) inside Box 2.) This difference is because in a thin film sample the magnetic precession is strongly elliptical, and a given strength of torque in the out-of-plane direction will excite a different amplitude of precession compared to the same strength of an in-plane torque.\\

\romannum{4}. Another error we have sometimes encountered involves how to measure $V^\text{mix}_\text{XX}$ (the d.c.\ mixing voltage) using a lock-in amplifier. If an ST-FMR measurement is conducted with a sinusoidal 100\% amplitude modulation, the amplitude of the driving rf current follows $I(t) = I_0(1+\sin(\omega_\text{m}t))$, where $I_0$ is the amplitude of the original driving current and $\omega_\text{m}$ is the modulation frequency. The first-harmonic component of the rf power is then expressed as: $P(t)\propto I(t)^2\propto 2I_0^2\sin(\omega_\text{m}t)$.
In the small angle precession approximation, the mixing voltage is proportional the applied power, and a lock-in output is the rms amplitude of the oscillations in its input. Therefore, $V_\text{lock-in}=\sqrt{2}V^\text{mix}_\text{XX}$\cite{karimeddiny2021}. \\

\romannum{5}. If utilizing the transverse + longitudinal ST-FMR technique of ref.~\cite{karimeddiny2020transverse} to correct for the spin-pumping artifact, care should be taken to use the same configuration of Hall leads to measure both the Hall resistance and the Hall ST-FMR signals. We provide a discussion of the measurement configuration and the artifact correction in Supplementary Information section \Romannum{2}.\\

\noindent\textbf{B. For second-harmonic Hall measurements:}\\

\romannum{1}. To minimize magnonic effects, choose a ferromagnet with large saturation magnetization and reasonably-large damping, and perform measurements at temperatures well below its Curie temperature.  \\

\romannum{2}. The out-of-plane damping-like signal is proportional to $R_\text{PHE}$, so avoid using a ferromagnet for which this is small, such as CoFeB\cite{he2022spin}.  The conventional in-plane damping-like signal is proportional to $R_\text{AHE}$, so if this is the term of primary interest, avoid materials with small anomalous Hall effects such as Ni$_{80}$Fe$_{20}$ and yttrium iron garnet.\\

\romannum{3}. Check whether magnon creation and annihilation contribute significantly to the signals using the method of No\"el \textit{et al.}\cite{noel2025nonlinear}, and if necessary attempt to correct for the magnetic effects as discussed in ref.~\cite{noel2025estimation}. Alternatively, if the unconventional damping-like torques are not too strong, perform the second-harmonic Hall measurements with the magnetic field rotated in the X-Z plane rather than within the sample plane\cite{xuAlternativeHH}.\\

\bigskip
\noindent \textbf{\Romannum{7}. Materials classes capable of generating out-of-plane damping-like torque}
\\

\noindent\textbf{A. Materials with low crystal symmetry}\\

\begin{figure}[H]
    	\centering
    	\includegraphics[width=1\textwidth]{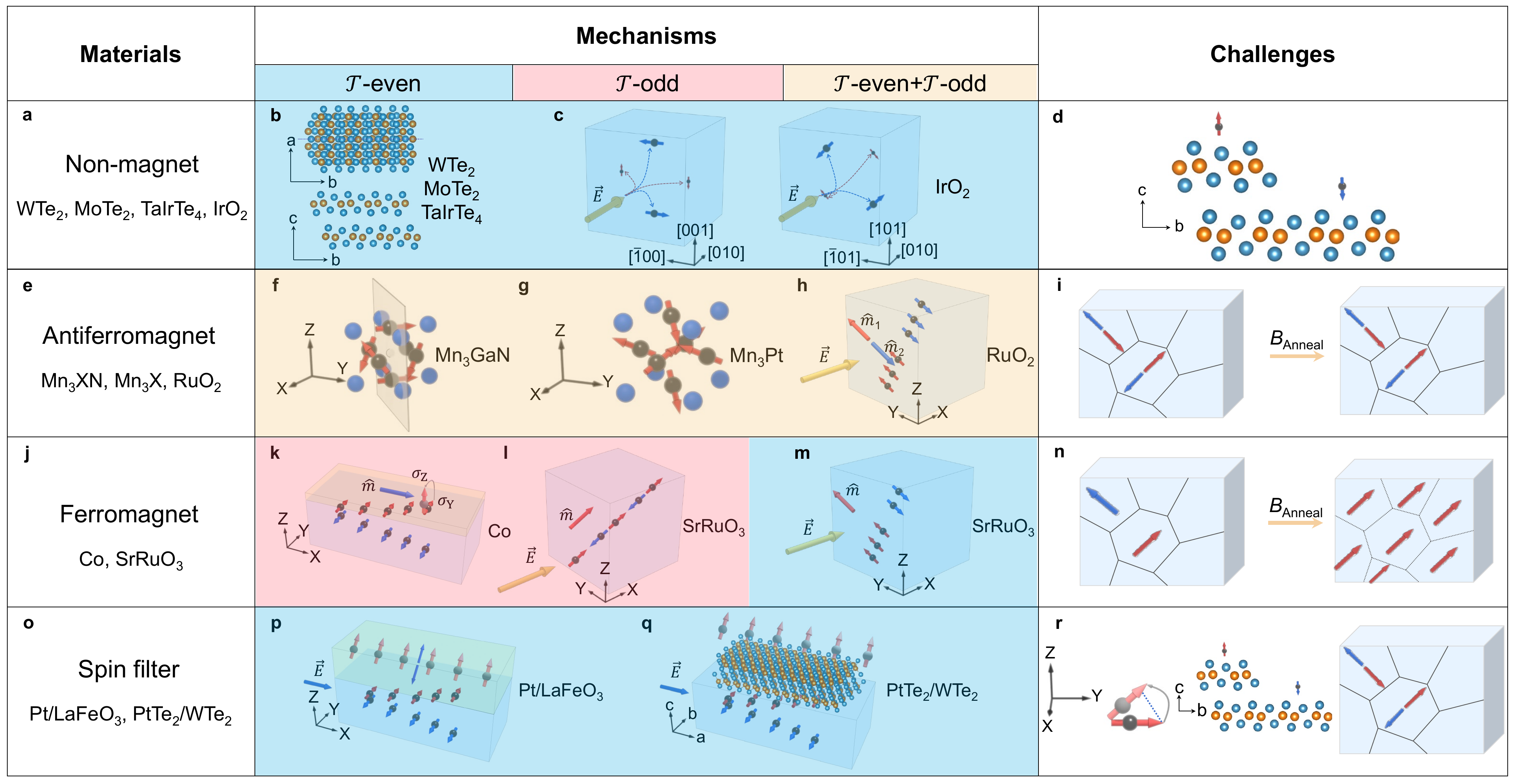}
    	\caption{\textbf{Pictorial summary of materials, mechanisms and challenges for producing $\tau_{\rm DL}^{\rm Z}$. a,} Examples of materials with low crystal symmetry which produce $\tau_{\rm DL}^{\rm Z}$. 
        \textbf{b, }Structure of the low-symmetry material WTe$_2$. 
        \textbf{c, }Depiction of electrically-generated spin currents in an anisotropic material when grown along a symmetry axis and away from a symmetry axis. \textbf{d, }The key challenge for producing $\tau_{\rm DL}^{\rm Z}$ with low-crystal-symmetry materials -- contributions from different structural domains can cancel. \textbf{e, }Examples of materials with antiferromagnetic ordering which produce $\tau_{\rm DL}^{\rm Z}$. 
        \textbf{f, }Structure of the noncollinear antiferromagnet Mn$_3$GaN for which the only allowed symmetry operation is a mirror plane along [1$\bar{1}$0]. 
        \textbf{g, }Magnetic spin Hall effect in non-collinear antiferromagnets. 
        \textbf{h, } Depiction of tilted spin currents generated by momentum-dependent spin splitting in an altermagnet. 
        \textbf{i, }The key challenge for producing $\tau_{\rm DL}^{\rm Z}$ with antiferromagnets is cancellation of contributions from differently-oriented magnetic domains. 
        \textbf{j, }Examples of ferromagnets for producing $\tau_{\rm DL}^{\rm Z}$. 
        \textbf{k, }Depiction of the spin rotation mechanism. 
        \textbf{l, }Depiction of the tilted spin current produced by the SPHE in a ferromagnet, where the magnetic moment is canted from $\rm Z$ towards $\rm X$ direction. 
        \textbf{m, }Depiction of the tilted spin current produced by the SAHE in a ferromagnet, where the magnetic moment is canted away from $\rm Z$ towards $\rm Y$ direction. 
        \textbf{n, }The challenge of producing $\tau_{\rm DL}^{\rm Z}$ with ferromagnets is cancellation of contributions from different magnetic domains; this might be addressed using an initializing magnetic field. 
        \textbf{o, }Candidate heterostructures for producing $\tau_{\rm DL}^{\rm Z}$ by spin filtering or spin relaxation. 
        \textbf{p, }Depiction of the spin filtering effect with an antiferromagnet to produce $\tau_{\rm DL}^{\rm Z}$. 
        \textbf{q, }Depiction of the spin filtering effect with a low symmetry crystal to produce $\tau_{\rm DL}^{\rm Z}$. 
        \textbf{r, }Key challenges for producing $\tau_{\rm DL}^{\rm Z}$ with spin filtering effect: Spin-orbit torque efficiency reduction by spin to spin conversion; potential cancellation induced by both structural and magnetic domains.     
        }. 
    	\label{fig:5}
\end{figure}

The first class of spin-source materials for which an out-of-plane component of damping-like torque was measured are materials with low crystal symmetry (Fig.~\ref{fig:5}a)\cite{macneill2017control,liu2023field,patton2023symmetry,patton2025crystallographic}. Several van der Waals semimetals including WTe$_2$\cite{macneill2017control}, MoTe$_2$\cite{stiehl2019layer}, and TaIrTe$_4$\cite{liu2023field} naturally possess crystal structures with sufficiently low symmetry to allow for out-of-plane damping-like torques (Fig.~\ref{fig:5}b).  For example, for samples made with WTe$_2$, there is only one mirror plane, $\mathcal{M}_{\rm bc}$ (defined by the b and c axes of the crystal), and an out-of-plane damping-like torque is produced when the charge current is applied along the a axis in the sample plane.  For these measurements it is important that the van der Waals materials can be isolated as well-ordered single crystals using exfoliation with perfect surfaces over micron-scale areas, without even any monolayer steps. With WTe$_2$, an extra monolayer at the interface with the magnetic free layer causes a sign change for the out-of-plane torque because adjacent layers in WTe$_2$ have crystal lattices rotated by 180$^\circ$\cite{macneill2017thickness}.  Samples containing monolayer steps can therefore produce canceling contributions to the net out-of-plane torque.\\

Out-of-plane damping-like torque can also be generated using materials with somewhat higher symmetry (e.g., tetragonal or orthorhombic), if the spin Hall tensor is anisotropic and the material is grown in a thin film oriented away from a symmetry axis\cite{patton2023symmetry,patton2025crystallographic}. For example, suppose that in a tetragonal spin-source material with electric field applied in the [010] direction the spin current flowing in the [100] direction is different than in the [001] direction.  (That is, in Fig.~\ref{fig:5}c the spin current colored red has a different magnitude than the spin current colored blue.) If the spin-source material is grown with the high-symmetry (001) orientation as shown in Fig.~\ref{fig:5}c (left), the damping-like torque will still be oriented strictly within the sample plane because only the spin current flowing out of plane can reach the interface and exert a torque on the adjacent magnet. However, if the same spin-source material is grown with a (101) orientation (Fig.~\ref{fig:5}c (right)), away from a high-symmetry direction, both the [001] and [100] crystal axes are rotated to have components perpendicular to the sample plane and so do the orientations of the spins flowing along the [001] and [100] directions. Since these two spin currents have different amplitudes, the total spin current calculated from the sum of these spin currents can have a non-zero component of out-of-plane spin flow with out-of-plane spin polarization, thus giving rise to an out-of-plane damping-like torque to an adjacent magnet.  This mechanism has been demonstrated using IrO$_2$ as the spin-source material\cite{patton2023symmetry,patton2025crystallographic}. \\

\noindent\textbf{B. Antiferromagnets}\\

In contrast to non-magnetic materials for which electrically-generated spin currents and associated torques must arise entirely from spin-orbit interactions, in magnetic materials -- either antiferromagnets or ferromagnets -- either exchange interactions or spin-orbit interactions (or their combination) can contribute.  This is intriguing because exchange interactions are generally stronger in magnitude than spin-orbit interactions. \\

Contributions to the spin currents and spin torques generated by antiferromagnetic ordering in a spin-source material can be classified into two groups, as either even or odd under the time-reversal operation ($\mathcal{T}$)\footnote{Spin torques arising entirely from spin-orbit interactions are naturally $\mathcal{T}$-even.}.  A $\mathcal{T}$-even case occurs when the combination of antiferromagnetic ordering and spin-orbit coupling leads to changes in the electronic bandstructure compared to the non-magnetic state, and these changes are invariant with respect to time reversal.  In this case the onset of antiferromagnetic order can lower the overall symmetry compared to the non-magnetic state, enabling the existence of unconventional components of spin-orbit torque upon magnetic ordering\cite{nan2020controlling,chen2021observation}.  For instance, for the spin-structure of a [001]-oriented thin film of cubic Mn$_3$GaN (Fig.~\ref{fig:5}f), there is only one mirror plane parallel to the (1$\bar{1}$0) plane and no two-fold rotational symmetry. 
This allows for a $\mathcal{T}$-even out-of-plane damping-like torque when a current is applied along [1$\bar{1}$0] direction\cite{nan2020controlling}\footnote{While the analysis in ref.\cite{nan2020controlling} supports the existence $\mathcal{T}$-even torques, Mn$_3$GaN might also generate $\mathcal{T}$-odd torques.}.\\

In addition, exchange interactions can generate $\mathcal{T}$-odd damping-like torques in antiferromagnets, but only antiferromagnets which lack both $\mathcal{PT}$ and $\mathcal{U}t$ symmetries, where $\mathcal{P}$ is inversion, $\mathcal{U}$ is spin flip, and $t$ is translation by half a unit cell. A $\mathcal{T}$-odd spin current generated by an antiferromagnet is known as the magnetic spin Hall effect\cite{kimata2019magnetic}.  If one or both of $\mathcal{PT}$ and $\mathcal{U}t$ symmetries is preserved, then the electronic band structure everywhere in momentum space remains Kramers-degenerate, while if both  are broken then the electronic states in momentum space can be spin-split\cite{hayami2019momentum,yuan2020giant,yuan2021}.  Depending on the direction of the applied electric field relative to the crystal axes, $\mathcal{T}$-odd spin currents in a spin-split bandstructure can flow both parallel and transverse to an applied electric field.  \\

Out-of-plane damping-like torques (both $\mathcal{T}$-even and $\mathcal{T}$-odd components) can occur in either non-collinear\cite{nan2020controlling,cao2023anomalous,pu2024field} or collinear\cite{bose2022tilted} antiferromagnets. Noncollinear antiferromagnets that can produce substantial out-of-plane damping-like torques include the families Mn$_3$XN (with X = Ga or Sn) and Mn$_3$X (with X = Pt, Ir, Sn, Ge, or Ga).\\

Collinear antiferromagnets which possess momentum-dependent spin splitting have been termed altermagnets\cite{hayami2019momentum,vsmejkal2020crystal,yuan2020giant,mazin2021prediction,vsmejkal2022emerging,vsmejkal2022beyond}.  
In altermagnets, the breaking of $\mathcal{PT}$ and $\mathcal{U}t$ symmetries results not from the orientations of the antiferromagnetic moments, but from a difference in the positioning of the non-magnetic atoms neighboring to each of the spin sublattices -- the environment of neighboring atoms about sublattice 1 is rotated relative to the environment of neighboring atoms about sublattice 2.  The result is that the electronic states are spin-polarized parallel to the N\'eel vector (and that component of spin remains a good quantum number), but the Fermi surfaces for spin-up and spin-down electrons are distinct, non-spherical, and rotated relative to each other in momentum space\cite{hayami2019momentum,yuan2020giant,vsmejkal2020crystal,mazin2021prediction,vsmejkal2022emerging}. Because the electronic states have spin parallel to the N\'eel vector but the spin-splitting does not result in spin-current flow parallel to the N\'eel vector, in order to generate an out-of-plane damping-like torque using an altermagnetic spin-source layer the layer must be grown with its N\'eel vector partially canted relative to the sample plane (neither fully in-plane nor perpendicular) (Fig.~\ref{fig:5}h).  Spin-current generation based on momentum-dependent spin splitting was predicted for RuO$_2$ \cite{gonzalez2021efficient}, and out-of-plane damping-like torque was subsequently measured in (101)-oriented RuO$_2$ \cite{bose2022tilted,karube2022observation}.  Whether this torque is due to momentum-dependent spin splitting associated with altermagnetism (rather than simply due to spin-orbit coupling within the anisotropic rutile crystal structure of RuO$_2$) remains uncertain, in part because whether or not RuO$_2$ is even an antiferromagnet remains uncertain\cite{berlijn2017itinerant,zhu2019anomalous,lin2024observation,hiraishi2024nonmagnetic,liu2024absence,kessler2024absence,kiefer2025crystal,song2025absence,osumi2026spin,jeong2026altermagnetic}. 
Evidence for the altermagnetic mechanism in ref.~\cite{bose2022tilted} includes that the vector orientation of the damping-like torque is closely aligned with the N\'eel vector and the temperature dependence of the out-of-plane damping-like torque points to a $\mathcal{T}$-odd mechanism\cite{bose2022tilted}.  However, a definitive test for a $\mathcal{T}$-odd mechanism would require testing whether the orientation of the torque is reversed when the N\'eel vector is reversed within a device, and this is difficult to accomplish in RuO$_2$ because the N\'eel vector (if it exists) does not couple to an applied magnetic field.\\

\noindent\textbf{C. Ferromagnets}\\

Several mechanisms have been proposed and experimentally verified for how ferromagnets can be used to produce spin currents with a spin polarization direction controllable by changing the magnetization direction\cite{amin2019intrinsic,davidson2020perspectives}, including spin rotation or spin swapping of spin currents incident at a magnetic interface\cite{saidaoui2016spin,humphries2017observation,baek2018spin,amin2018interface,yang2024field}, the spin planar Hall effect (SPHE)\cite{taniguchi2015spin,safranski2019spin,safranski2020planar}, and the spin anomalous Hall effect (SAHE)\cite{taniguchi2015spin,iihama2018spin,seki2019large,montoya2025anomalous}. If a ferromagnet possesses crystalline anisotropy, an electric field applied away from a high-symmetry direction should also produce unconventional spin torques due simply to the anisotropy of the spin-split Fermi surface, analogous to an antiferromagnet with an anisotropic spin-split Fermi surface\cite{gonzalez2021efficient}.\\ 

In the spin rotation mechanism, an incident spin current with spin moment in the $\rm Y$ direction can rotate in an exchange field oriented in the $\rm X$ direction to produce a component of spin moment in the $\rm Z$ direction (Fig.~\ref{fig:5}k). Out-of-plane damping-like torque due to spin rotation has been inferred from field-free domain-wall-mediated switching and hysteresis loop measurements, but not yet measured quantitatively. The quantitative measurements of this mechanism that have been performed utilized a PMA magnet for the spin rotation so that an incident spin current with spin moment in the $\rm Y$ direction rotated to produce a component of spin moment in the $\rm X$ direction, giving an unconventional in-plane damping-like torque. These  measurements so far suggest that this mechanism is weak, with an unconventional efficiency of $\xi_{\rm DL}^{\rm X}$=0.0048\cite{humphries2017observation} or $\sim0.003$\cite{wang2020detection}. \\

Both the SPHE and SAHE give rise to spin currents with spin polarization aligned with the magnetization of the spin-source layer ($\hat{m}$), but with different flow directions: for the SPHE the flow is parallel to $\hat{m}(\hat{m}\cdot\vec{E})$ (Fig.~\ref{fig:5}l) and for the SAHE the flow is along $\hat{m}\times\vec{E}$ direction (Fig.~\ref{fig:5}m). It follows that to generate an out-of-plane damping-like torque using either the SPHE or the SAHE requires that the magnetization direction of the spin-current source be canted relative to the sample plane -- neither fully in-plane or perpendicularly magnetized.  Taking into account both the spin orientation and flow direction, the out-of-plane damping-like torque due to the SPHE should be proportional to $m_\text{X} m^2_\text{Z}$ so that it is $\mathcal{T}$-odd, and that due to the SAHE should be proportional to $m_\text{Y} m_\text{Z}$ so that it is $\mathcal{T}$-even.  The initial experiments to investigate the efficiencies of the SPHE and SAHE mechanisms have been performed using  in-plane-magnetized spin-source layers\cite{iihama2018spin,seki2019large} (so that the resulting damping-like torque was also strictly in-plane) or by using large canted saturating magnetic fields to control the magnetization of the spin-source layer \cite{safranski2019spin,safranski2020planar,montoya2025anomalous}. \\
 
To make out-of-plane damping-like torque due to either the SPHE or the SAHE useful for practical applications, the partially canted magnetic configuration must be stabilized in the absence of an applied magnetic field. Thin film samples governed entirely by shape anisotropy and first-order interfacial anisotropy have either purely in-plane or perpendicular magnetic anisotropy, so to have a stable partially canted magnetic configuration requires use of magneto-crystalline anisotropy.  One way to achieve a canted configuration is to deposit an oxide ferromagnet such as SrRuO$_3$ onto a vicinal substrate cut at a shallow angle; nucleation of growth from step edges in the substrate can stabilize primarily one domain variant with a tilted magnetization.  Using this approach, out-of-plane damping-like torque has been demonstrated below the Curie temperature of SrRuO$_3$ ($\sim$ 160 K), with contributions from both the SPHE and SAHE\cite{huang2025control}.  
Calculations suggest that other magnetic material with strong magnetocrystalline anisotropy might be able to serve as even better spin sources, including \textit{L}1$_0$ FePt, FePd, or CoPt\cite{belashchenko2024exchange}. \\

\noindent\textbf{D. Heterostructures with spin filtering or spin relaxation}\\

One need not consider spin sources composed of just one material layer -- it is also possible to make composite spin sources from heterostructured bilayers in which one layer produces a strong spin current with a conventional orientation (in the Y direction) and then a spacer layer filters or rotates the orientation of the spin before it reaches a magnetic layer, so that it can exert an unconventional torque (Fig.~\ref{fig:5}p,q).  The spacer layer might in principle be any material with a symmetry that is sufficiently reduced to enable an out-of-plane damping-like torque, e.g.\ a low-symmetry crystalline material, an antiferromagnet, or a ferromagnet.
Proof-of-principle tests of this idea have been performed with a PtTe$_2$/WTe$_2$ bilayer in which PtTe$_2$ is the conventional spin source and WTe$_2$ is the spin-filter spacer layer\cite{wang2024field}, and with a Pt/LaFeO$_3$ bilayer in which Pt is the conventional spin source and a single-domain film of the antiferromagnetic insulator LaFeO$_3$ is the spacer\cite{chen2025pure}. An out-of-plane anti-damping torque from WTe$_2/$NiO bilayers has also been reported, interpreted as due to a perpendicularly-polarized component of spin current from WTe$_2$ that was transmitted via magnons through the antiferromagnetic insulator NiO\cite{wang2024field}.\\

\noindent\textbf{E. Quantitative comparison of experimental results}\\

In Table \ref{table 1} we compile the quantitative measurements announced to date for the spin-source materials for which non-zero values of out-of-plane damping-like torque have been reported, as well as the measurement technique used, broken down by the source of symmetry breaking: crystals with low crystal symmetry, antiferromagnets, ferromagnets, and spin-filter samples. We present this table with some important caveats, however.  The majority of measurements reported used ST-FMR with the ``S/A'' approach introduced in ref.\ \cite{liu2011spin} to estimate the rf current amplitude; as noted above, if the conventional Y component of the field-like spin-orbit torque is not negligible compared to the Oersted torque then this approach will produce significant errors. A direct calibration of the rf current should be performed instead\cite{patton2023symmetry,huang2025control,liang2026unconventional}. \\

The column ``Orient. Dep.?'' in Table \ref{table 1} indicates whether an in-plane crystallographic orientation dependence has been established to reconstruct the torque symmetry of the out-of-plane damping-like torque. We consider such orientation-dependent measurements a crucial test for assessing whether unconventional damping-like torques reflect intrinsic symmetry-allowed physics rather than experimental artifacts. \\

Several experiments reported evidence for the existence of non-zero values of $\xi^\text{Z}_\text{DL}$ based on a nonlinear-in-current loop shift or field-free-switching measurements on micron scale samples; in the Table we wish to credit these works but since these techniques do not provide quantitative values we leave the quantitative values for these entries blank.  Blank values are also listed for a few other papers which reported qualitative but not quantitative results.  \\

For completeness, Table \ref{table 1} includes experimental reports even for samples which by symmetry we expect that $\xi^\text{Z}_\text{DL}$ actually must be zero. We are puzzled, for example, by the reports of nonzero values of $\xi^\text{Z}_\text{DL}$ from non-epitaxial deposited films of WTe$_2$ in micrometers-scale samples\cite{shi2021observation,wang2023room,shi2024field}. We would expect the contributions to $\xi^\text{Z}_\text{DL}$ from randomly-oriented grains of WTe$_2$ in a polycrystalline film to cancel.
Even a single-domain sample of WTe$_2$ with surface roughness should have a strongly diminished value of $\xi^\text{Z}_\text{DL}$ because the sign of $\xi^\text{Z}_\text{DL}$ reverses across a monolayer step in WTe$_2$ if the step is located at the interface with the magnet\cite{macneill2017thickness}.    \\

\noindent \textbf{\Romannum{8}. Outlook}\\

\begin{figure}[H]
    	\centering
    	\includegraphics[width=0.5\textwidth]{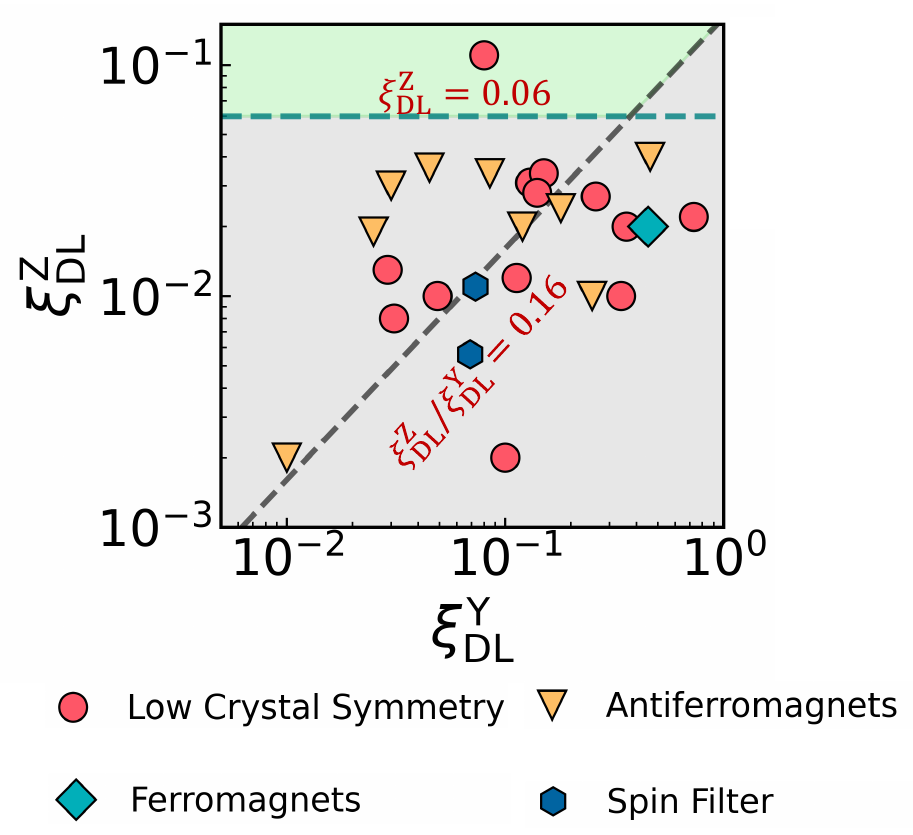}
    	\caption{\textbf{Survey of measurements of out-of-plane damping-like torque.} A summary of both in-plane and out-of-plane damping-like torque efficiencies for four classes of materials systems. The light green shaded area enclosed by the dashed lines ($\xi_{\text{DL}}^\text{Z}/\xi_{\text{DL}}^\text{Y}=0.16$ and $\xi_{\text{DL}}^\text{Z}=0.06$) indicates the regime where the write energy of 3T SOT-MRAM becomes competitive compared to STT-MRAM. 
        }. 
    	\label{fig:6}
\end{figure}

We discussed in Section III above the specifications needed in order that out-of-plane damping-like torque might contribute to useful technologies: e.g., in order for the write energy of a 3T SOT-MRAM device to become competitive compared to state-of-the-art existing STT-MRAM one needs $\xi^\text{Z}_\text{DL}> 0.06$ and also $\xi^\text{Z}_\text{DL}/\xi^\text{Y}_\text{DL}> 0.16$\cite{liao2020spin,shazon_TED}.
Fig. \ref{fig:6} shows how the experimental results to date for different spin-source materials compare to these initial goals. The research field has reached the stage where several different materials are approaching the figures of merit for usefulness, with one report surpassing them, for the van der Waals semimetal TaIrTe$_4$\cite{bainsla2024large}.  As noted in Table \ref{table 1}, however, other studies of TaIrTe$_4$ report lower values of $\xi^\text{Z}_\text{DL}$.  \\

Whether a new out-of-plane spin-source material will be capable of enabling a useful memory technology depends not only on the torque efficiencies, but also on its potential for manufacturability, scalability, and compatibility with current semiconductor technology.  The different classes of spin-source materials that we have discussed each have different challenges in this regard, as depicted in the ``Challenges'' column of Fig.~\ref{fig:5}.
Materials with low crystal symmetry will generally require samples made from single-crystal layers with minimal structural disorder, as differently-oriented crystalline domains can give canceling contributions to the out-of-plane damping-like torque. The situation is even worse for the van der Waals materials WTe$_2$ and TaIrTe$_4$ in which the orientation of the broken symmetry that enables the out-of-plane damping-like torque reverses between adjacent crystal layers\cite{macneill2017thickness}; therefore, even single-monolayer steps at the interface with the magnetic layer in single-crystal samples  will produce regions with canceling contributions to $\xi^\text{Z}_\text{DL}$ (Fig.~\ref{fig:5}d).  CMOS-compatible thin film growth techniques such as magnetron sputtering, chemical vapor deposition, and atomic layer deposition are not capable of producing single-crystalline low-crystal symmetry layers, much less perfectly-smooth layers without any atomic steps.  Antiferromagnetic spin source materials will experience similar cancellations between the contributions from different antiferromagnetic domains, and it will also be a difficult challenge to fabricate devices from single-domain antiferromagnets (Fig.~\ref{fig:5}i).  \\ 

We are most optimistic about spin-source materials that do not require epitaxial growth, and where the net orientation of the symmetry breaking that allows for a non-zero value of $\xi^\text{Z}_\text{DL}$ can be set by an initializing field after deposition, to minimize the extent of cancellation between domains.  Ferromagnets with strong magneto-crystalline anisotropy and antiferromagnets which possess a small net magnetic moment because of a Dzyaloshinskii-Moriya interaction might fall in this category.  Fig. \ref{fig:5}n illustrates this idea in cartoon form.  Among the mechanisms available for generating spin currents from ferromagnets, the SAHE and the SPHE have shown potential for generating useful out-of-plane damping-like torque efficiencies, if reliable ways can be developed to stabilize canted magnetic configurations in appropriate source materials. \\

We will close by noting that to identify materials that are both manufacturable at scale and that can drive efficient  anti-damping switching of nanomagnets with PMA is one of the most important open challenges in the field of spintronics. Additional materials innovation will be needed to develop out-of-plane damping-like torque to the point where it can be useful for new magnetic technologies competitive with the existing memory technologies of STT-MRAM and SRAM. To assess whether a new out-of-plane spin-source material is promising for energy-efficient SOT-MRAM technology, rigorous analyses of both torque symmetry and torque strength are essential.  Quantitative measurements are necessary, and should be interpreted carefully to extract reliable torque efficiencies, with particular attention paid to avoiding potential artifacts from spin pumping, current-induced heating, magnon creation and annihilation, inaccurate charge-current calibration, and related effects. The existence of unconventional torques should furthermore be verified through their dependence on the direction of applied electric field relative to crystallographic orientation and/or magnetic order, as such torques necessarily arise from symmetry breaking.  

\pagebreak

\begin{center}
\begin{longtable}{|c|c|c|c|c|c|c|c|} 
 \hline
 Material & Fab. Tech. & $\xi_{\rm DL}^{\rm Z}$ & $\xi_{\rm DL}^{\rm Y}$ &Measure. Tech.&\makecell{Orient. \\Dep.?}&\makecell{Tech. \\Relevance}& ref.\\ 
 \hline
 \multicolumn{8}{|c|}{Crystals with low crystal symmetry}\\
 \hline
\multirow{5}{*}{WTe$_2$} 
& Mech. Exfo. & 0.013 & 0.029  & \makecell{ST-FMR with \\curr. calib.}& \textbullet & \multirow{5}{*}{\makecell{Air\\sensitive}}& \cite{macneill2017control} \\ \cline{2-6}\cline{8-8}
& CVD& 0.031& 0.13& \makecell{ST-FMR with \\curr. calib.}
& --& & \cite{shi2021observation} \\ \cline{2-6}\cline{8-8}
& CVD& 0.02& 0.36& SHH& --& & \cite{wang2023room} \\ \cline{2-6}\cline{8-8}
& MBE& 0.034& 0.15& ST-FMR with S/A& \textbullet& &\cite{wang2024field} \\ \cline{2-6}\cline{8-8}
& MBE& 0.028& 0.14& ST-FMR with S/A& \textbullet& & \cite{shi2024field} \\ \cline{2-8}
 \hline
MoTe$_2$&Mech. Exfo.&0.008&0.031&\makecell{ST-FMR with \\curr. calib.}&\textbullet&\makecell{Air\\sensitive}&\cite{stiehl2019layer}\\
\hline
\makecell{(114)\\MnPd$_3$}&MS&0.01&0.34&SHH&--&\makecell{Struct.\\Domain}&\cite{dc2023observation}\\
\hline
\multirow{4}{*}{TaIrTe$_4$}
&Mech. Exfo. &0.012$^*$&0.113&ST-FMR with S/A&--&\multirow{4}{*}{\makecell{Air\\sensitive}}&\cite{liu2023field}\\\cline{2-6} \cline{8-8}
&Mech. Exfo. &$\sim 0$&0.05&ST-FMR with S/A&--& &\cite{zhang2023room}\\\cline{2-6} \cline{8-8}
&Mech. Exfo. &0.11&0.08&\makecell{ST-FMR with S/A\\SHH}&--& &\cite{bainsla2024large}\\ \cline{2-6}\cline{8-8}
&Mech. Exfo. &--&--&\makecell{Loop shift\\Field-free switching}&--& &\cite{zhang2024robust} \\ \cline{2-6}\cline{8-8}
\hline
RuO$_2$&MBE&0.01&0.049&\makecell{ST-FMR with \\curr. calib.\\SHH}&\textbullet&\makecell{Scalability}&\cite{bose2022tilted}\\
\hline
IrO$_2$&PLD&0.002&0.1&\makecell{ST-FMR with \\curr. calib.}&\textbullet&Scalability&\cite{patton2025crystallographic}\\
\hline
Ni$_4$W&MS&0.022&0.73&SHH&\textbullet&\makecell{Struct.\\Domain}&\cite{yang2025large}\\
\hline
WO$_2$&MS&0.027&0.26&ST-FMR with S/A&\textbullet&\makecell{Struct.\\Domain}&\cite{song2025anomalous}\\
\hline
CaRuO$_3$&PLD&0.0086&1.4&ST-FMR with S/A&--&Scalability&\cite{han2024generation}\\
\hline
Mn$_2$Au&MS&--&--&Field-free switching&--&\makecell{Mag. \\Domain}&\cite{chen2021observation}\\
\hline
 \multicolumn{8}{|c|}{Antiferromagnets}\\
\hline
Mn$_3$GaN&MS&0.019&0.025&\makecell{ST-FMR with \\curr. calib.}&--&\makecell{Mag.\\Domain}&\cite{nan2020controlling}\\
\hline
Mn$_3$SnN&MS&0.003&--&ST-FMR with S/A&--&\makecell{Mag.\\Domain}&\cite{you2021cluster}\\
\hline
\multirow{5}{*}{Mn$_3$Sn}
&Bridgeman&$\sim 0$&--&ST-FMR&--&\multirow{5}{*}{\makecell{Mag.\\domain}}&\cite{kondou2021giant}\\\cline{2-6} \cline{8-8}
&MS&--&--&\makecell{Loop shift\\Field-free switching}&--& &\cite{hu2022efficient}\\ \cline{2-6}\cline{8-8}
&MS&0.002$^*$&0.01$^*$&ST-FMR with S/A&--& &\cite{hazra2023generation}\\\cline{2-6} \cline{8-8}
&MS&0.03&$\sim 0$&ST-FMR with S/A&--& &\cite{meng2024field}\\ \cline{2-6}\cline{8-8}
&MS&0.03&0.03&ST-FMR with S/A&--& &\cite{meng2025correlation}\\\cline{2-6} \cline{8-8}
\hline
\makecell{(11$\bar{2}$0)\\Mn$_3$Sn}&MS&0.027&--&ST-FMR with S/A&--&\makecell{Mag.\\domain}&\cite{lu2026enhancement}\\ 
\hline
Mn$_3$Ga&MBE&--&--&\makecell{Loop shift\\Field-free switching}&\textbullet&\makecell{Mag.\\domain}&\cite{han2023field}\\ 
\hline
Mn$_3$Ge&MS&0.01&0.25&ST-FMR with S/A&--&\makecell{Mag.\\domain}&\cite{cao2025unconventional}\\
\hline
\multirow{4}{*}{Mn$_3$Pt}
&MS&0.034$^*$&0.085&ST-FMR with S/A&--&\multirow{4}{*}{\makecell{Mag.\\domain}}&\cite{bai2021control}\\ \cline{2-6}\cline{8-8}
&MS&$\sim 0$&--&ST-FMR&--& &\cite{yu2022observation} \\ \cline{2-6}\cline{8-8}
&MS&0.036&0.045&ST-FMR with S/A&\textbullet& &\cite{cao2023anomalous}\\ \cline{2-6}\cline{8-8}
&MS&0.04&0.46&SHH&--& &\cite{novakov2023composite}\\ \cline{2-6}\cline{8-8}
\hline
\multirow{2}{*}{Mn$_3$Ir}
&MS&0.02&0.12&ST-FMR with S/A&\textbullet&\multirow{2}{*}{\makecell{Mag.\\domain}}&\cite{pu2024field}\\\cline{2-6}\cline{8-8}
&MS&0.024&0.18&ST-FMR with S/A&--& &\cite{karim2024observation}\\\cline{2-6}\cline{8-8}
\hline
 \multicolumn{8}{|c|}{Ferromagnets}\\
\hline
\makecell{IP FM/Ti/\\CoFeB/MgO}&MS&--&--&\makecell{Loop shift\\Field-free switching}&--&\makecell{\textit{B} field\\sensitive}&\cite{baek2018spin}\\
\hline
IP epi-Co&MS&--&--&Field-free switching&\textbullet&\makecell{\textit{B} field\\sensitive}&\cite{ryu2022}\\
\hline
\makecell{canted\\ SrRuO$_3$}&PLD&0.013&0.36&\makecell{ST-FMR with \\curr. calib.}&\textbullet&\makecell{Cryogenic}&\cite{huang2025control}\\
\hline
 \multicolumn{8}{|c|}{Heterostructures with spin filtering or spin relaxation}\\
\hline
WTe$_2$/NiO&MBE&0.011&0.073&ST-FMR with S/A&\textbullet&\makecell{Air\\sensitive}&\cite{wang2024deterministic}\\
\hline
Pt/LaFeO$_3$&PLD&0.0056&0.069&ST-FMR with S/A&\textbullet&scalability&\cite{chen2025pure}\\
\hline
PtTe$_2$/WTe$_2$&MBE&\makecell{Thick.\\dep.}&\makecell{Thick.\\dep.}&\makecell{ST-FMR with S/A\\Field-free switching}&\textbullet&\makecell{Air\\sensitive}&\cite{wang2024field}\\
\hline
\caption{A list of materials for which measurements of non-zero out-of-plane damping-like torque have been reported. $^*$Our calculation based on the data in the cited paper. Mech. Exfo.: Mechanical Exfoliation; curr. calib.: current calibration performed. CVD: Chemical Vapor Deposition; MBE: Molecular Beam Epitaxy; PLD: Pulsed Laser Deposition; MS: Magnetron Sputtering; SHH: Second-harmonic Hall measurement. IP: in-plane. Orient. Dep.?: Were measurements performed as a function of the orientation of applied electric field relative to the crystal axes? \textbullet: Yes, orientation-dependent measurements were reported. 
Tech. Relevance: Technological relevance. This column highlights limitations specific to each material class.
Thick. Dep.: thickness-dependence measurements.}
\label{table 1}
\end{longtable}
\end{center}
\pagebreak

\noindent \textbf{Acknowledgments}\\

We thank Azad Naeemi, Ding-Fu Shao, and Evgeny Tsymbal for discussions.  X.H.\ and Y.L.\ were supported by the US Department of Energy (DOE),  Office of Basic Energy Sciences (BES) under award number DE-SC0017671.  X.H.\ led the writing of the review, and Y.L.\ provided significant assistance and discussion.  O.S.\ was supported in part by the Center for Energy Efficient Magnonics, an Energy Frontier Research Center funded by the U.S. Department of Energy, Office of Science, Basic Energy Sciences under Award DE-AC02-76SF00515, and in part by a New Frontier Award from the Arts \& Sciences College of Cornell University. He wrote the supplementary section on sign conventions for transverse + longitudinal ST-FMR.  D.C.R.\ and some research expenses were supported by SUPREME, one of seven centers in JUMP 2.0, a Semiconductor Research Corporation program sponsored by DARPA.  D.C.R.\ provided supervision and editorial suggestions.  Device fabrication was performed using the shared facilities of the Cornell NanoScale Facility, a member of the National Nanotechnology Coordinated Infrastructure (supported by the NSF via grant NNCI-2025233) and the facilities of the Cornell Center for Materials research.

\pagebreak
\bibliographystyle{naturemag}
\bibliography{bibliography}
\pagebreak
\clearpage

\begin{center}
    {\Large\bfseries Supplementary Information\par}
    \vspace{0.8em}
    {\large\bfseries Materials and Metrology for Out-of-Plane Damping-Like Spin-Orbit Torque\par}
    \vspace{0.8em}
    {\large Xiaoxi Huang et al.\par}
\end{center}

\vspace{1.5em}
\setcounter{equation}{0}
\renewcommand{\theequation}{S\arabic{equation}}

\noindent \textbf{\Romannum{1}. Methods to evaluate the microwave current $I_\text{rf}$ for ST-FMR measurements}\\

We mentioned in the main text that in order to make ST-FMR measurements quantitative, it is important to determine the microwave current $I_\text{rf}$ by measurements, rather than relying on the ``S/A'' approach introduced in ref.~\cite{liu2011spin}.  This is because the ``S/A'' method assumes that the contribution to the field-like torque $\tau^0_\text{FL,Y}$ is due entirely to the Oersted field, and this can be in error because the spin-orbit-torque contribution to the field-like component can for some materials be comparable to or even larger than the Oersted contribution.  Here we discuss two general approaches: direct calibrations of $I_\text{rf}$, and studies using samples with different layer thicknesses to separate interfacial field-like SOTs from volume-acting Oersted torques.\\

\noindent\textbf{A. Estimate of damping-like SOT efficiency with direct current calibration}\\

As long as samples are made on high-resistivity substrates so that microwave absorption by the substrate is negligible, the microwave current flowing in the sample can be calibrated by using a vector network analyzer to measure the attenuation in the microwave lines and the complex reflection coefficient ($S_\text{11}$) at the sample. $I_\text{rf}$ (peak) can then be determined as:
\begin{equation}
    I_\text{rf} = \sqrt{\frac{ 2P_\text{delivered}}{Real(Z(\omega))}},
\end{equation}
where $P_\text{delivered} = P_\text{incident}(1-\left| S_\text{11} \right|^2)$, $Z(\omega) = Z_\text{0}(1+S_\text{11})/(1-S_\text{11})$ and $Z_\text{0}$ is the port reference impedance (50 $\Omega$). \\

Another alternative is to compare the heating-induced change in resistance from the microwave current to the heating-induced change from a d.c.\ current\cite{patton2023symmetry,huang2025control}.   \\

Once $I_\text{rf}$ is determined, the current density within the spin-source material, $J_\text{e}$, can be estimated from a parallel circuit model using the measured resistivities of the spin-source material and detecting ferromagnet.   If one defines $S_{\text{DL}}^\text{X}$, $S_{\text{DL}}^\text{Y}$, $A_{\text{DL}}^\text{Z}$, $A_{\text{FL}}^\text{X}$, $A_{\text{FL}}^\text{Y}$, $S_{\text{FL}}^\text{Z}$ as the amplitudes of the mixing voltages contributed by the X, Y, Z components of damping-like and field-like torques:
\begin{equation}
        V_\text{S} =  S^\text{X}_\text{DL} \sin2\varphi \sin\varphi  + S^\text{Y}_\text{DL} \sin2\varphi \cos\varphi + S^\text{Z}_\text{FL} \sin2\varphi, 
    \end{equation}
    \begin{equation}
        V_\text{A} =  A^\text{X}_\text{FL} \sin2\varphi \sin\varphi + A^\text{Y}_\text{FL} \sin2\varphi \cos\varphi + A^\text{Z}_\text{DL} \sin2\varphi .\\ 
    \end{equation}
then using Eq.~(6), Eq.~(B1), (B2), and (B6) in Box 2 in the main text, the final values of the damping-like torque efficiencies can then be written as\cite{huang2025control}:
\begin{equation}
    \xi_{\text{DL}}^\text{X} = S_{\text{DL}}^\text{X}\frac{eM_\text{s}t_{\text{FM}}}{\hbar J_\text{e}}\frac{4\Delta}{I_{\text{rf}}R_{\text{AMR}}}\frac{2B_0+\mu_0M_{\text{eff}}}{\sqrt{B_0(B_0+\mu_0M_{\text{eff}})}},
    \label{X_2} 
\end{equation}
\begin{equation}
    \xi_{\text{DL}}^\text{Y} = S_{\text{DL}}^\text{Y}\frac{eM_\text{s}t_{\text{FM}}}{\hbar J_\text{e}}\frac{4\Delta}{I_{\text{rf}}R_{\text{AMR}}}\frac{2B_0+\mu_0M_{\text{eff}}}{\sqrt{B_0(B_0+\mu_0M_{\text{eff}})}},
    \label{Y_2}
\end{equation}
\begin{equation}
    \xi_{\text{DL}}^\text{Z} = A_{\text{DL}}^\text{Z}\frac{eM_\text{s}t_{\text{FM}}}{\hbar J_\text{e}}\frac{4\Delta}{I_{\text{rf}}R_{\text{AMR}}}\frac{2B_0+\mu_0M_{\text{eff}}}{B_0+\mu_0M_{\text{eff}}}.
    \label{Z_2}
\end{equation}\\
Note that the part of the expression involving $B_0$ and $\mu_0 M_\text{eff}$ differs in the determination of $\xi^\text{Z}_\text{DL}$ (Eq.\ (\ref{Z_2})) compared to $\xi^\text{X}_\text{DL}$ and $\xi^\text{Y}_\text{DL}$ (Eqs.\ (\ref{X_2},\ref{Y_2})). This is because for elliptical precession of a thin-film sample a torque in the out-of-plane direction is more effective at exciting the precession compared to the same value of torque applied in the plane. (This follows from the $\omega_2/\omega$ term multiplying $\tau^0_\text{DL,Z}$ in Eq.\ (B6) in Box 2 in the main text. \\

\noindent\textbf{B. Estimate of SOT efficiency with field-like torque calibration}\\

Alternatively, the approach of ref.~\cite{liu2011spin} can be extended to account for the effects of field-like SOT by performing ST-FMR measurements on a series of samples with different layer thicknesses\cite{pai2015dependence}. To apply this technique, the layer whose thickness is varied must remain sufficiently thick so that the magnitudes of the interfacial SOTs do not change with its thickness, and the only remaining thickness dependence is the scaling of the Oersted torque. In principle, one can vary the thickness of either the detector ferromagnet or the spin-source layer, but varying the ferromagnet thickness is generally easier because for a strong ferromagnet the spin absorption length beyond which the field-like torque does not change with thickness tends to be shorter than the spin diffusion length in the spin-source layer. \\ 

It is useful to first define the intermediate quantities
\begin{equation}
        \xi_{\text{FMR}}^\text{X} = \frac{S_{\text{DL}}^\text{X}}{A_{\text{FL}}^\text{Y}} \frac{e \mu_0 M_\text{s} t_{\text{FM}} t_{\text{HM}}}{\hbar} \sqrt{1+\frac{\mu_0 M_{\text{eff}}}{B_0}},
        \label{X_1}
\end{equation}
\begin{equation}
        \xi_{\text{FMR}}^\text{Y} = \frac{S_{\text{DL}}^\text{Y}}{A_{\text{FL}}^\text{Y}} \frac{e \mu_0 M_\text{s} t_{\text{FM}} t_{\text{HM}}}{\hbar} \sqrt{1+\frac{\mu_0 M_{\text{eff}}}{B_0}},
        \label{Y_1}
\end{equation}
\begin{equation}
        \xi_{\text{FMR}}^\text{Z} = \frac{A_{\text{DL}}^\text{Z}}{A_{\text{FL}}^\text{Y}} \frac{e \mu_0 M_\text{s} t_{\text{FM}} t_{\text{HM}}}{\hbar}.
        \label{Z_1}
\end{equation}
Using Eq.~(7) and Eq.~(B5) in Box 2 in the main text, and the fact that the Oersted torque $\tau_{\text{Oe}}^0=\gamma\mu_0J_\text{e}t_{\text{HM}}/2$ by Ampere's law, the contributions to $A_{\text{FL}}^\text{Y}$ from both the field-like SOT and the Oersted torque can be written: 
\begin{equation}
A_{\text{FL}}^\text{Y} = \frac{I_{\text{rf}}R_{\text{AMR}}}{2\alpha\omega^+}\frac{\omega_2}{\omega} (\xi_{\text{FL}}^\text{Y}\frac{\mu_\text{B}J_\text{e}}{eM_\text{s}t_{\text{FM}}}+\frac{\gamma \mu_0 J_\text{e} t_{\text{HM}}}{2}).
\label{A}
\end{equation}
Combining Eq.~(6) and Eqs.~(B1), (B2), and (B6) in Box 2 from the main text with  Eqs.~(\ref{X_1})-(\ref{A}), we get:
\begin{equation}
    \frac{1}{\xi_{\text{FMR}}^\text{X}} = \frac{1}{\xi_{\text{DL}}^\text{X}}\Big(1+\frac{\hbar}{e}\frac{\xi_{\text{FL}}^\text{Y}}{\mu_0 M_\text{s} t_{\text{FM}}t_{\text{HM}}}\Big),
    \label{FMRX}
\end{equation}
\begin{equation}
    \frac{1}{\xi_{\text{FMR}}^\text{Y}} = \frac{1}{\xi_{\text{DL}}^\text{Y}}\Big(1+\frac{\hbar}{e}\frac{\xi_{\text{FL}}^\text{Y}}{\mu_0 M_\text{s} t_{\text{FM}}t_{\text{HM}}}\Big),
    \label{FMRY}
\end{equation}
\begin{equation}
    \frac{1}{\xi_{\text{FMR}}^\text{Z}} = \frac{1}{\xi_{\text{DL}}^\text{Z}}\Big(1+\frac{\hbar}{e}\frac{\xi_{\text{FL}}^\text{Y}}{\mu_0 M_\text{s} t_{\text{FM}}t_{\text{HM}}}\Big).
    \label{FMRZ}
\end{equation}
To determine SOT efficiencies $\xi_{\text{DL}}^\text{K}$, one can plot $\xi_{\text{FMR}}^{\text{K}}$ at a series of ferromagnet-layger thicknesses (or heavy-metal-layer thickness) and perform a linear fit of $1/\xi_{\text{FMR}}^\text{K}$ versus $1/t_{\text{FM}}$. The damping-like torque efficiency is the intercept as $1/t_{\text{FM}}\rightarrow0$, and the field-like torque efficiency $\xi^\text{Y}_\text{FL}$ can then be determined from the slope. \\

This technique is particularly useful when measuring samples on non-high-resistivity substrates, for which microwave absorption by the substrates makes it clumsy to directly calibrate $I_\text{rf}$ by the techniques discussed in the previous section. A caveat is that if the layer thicknesses are made too large, then contributions from spin pumping (see main text) can interfere with the fits, because they can cause the dependence of $\xi^\text{K}_\text{FMR}$ versus $1/t_{\text{FM}}$ to become nonlinear, at least for the $\xi^\text{Y}_\text{DL}$ contribution\cite{tapping2021}. The assumption that the interfacial spin torque does not depend on layer thickness can also be invalidated if orbital torques give a significant contribution.\\

\noindent \textbf{\Romannum{2}. Sign conventions for transverse + longitudinal ST-FMR}\\

The extraction of artifact voltages from ST-FMR measurements using the approach of ref.~\cite{karimeddiny2020transverse} depends on the sign convention used for measuring Hall voltages. A mistake in this sign convention can be difficult to detect because, in the absence of artifacts, this does not affect the extracted torque amplitudes. However, it can lead to an incorrect artifact correction.\\

\begin{figure}
    \centering
    \includegraphics[width=0.5\linewidth]{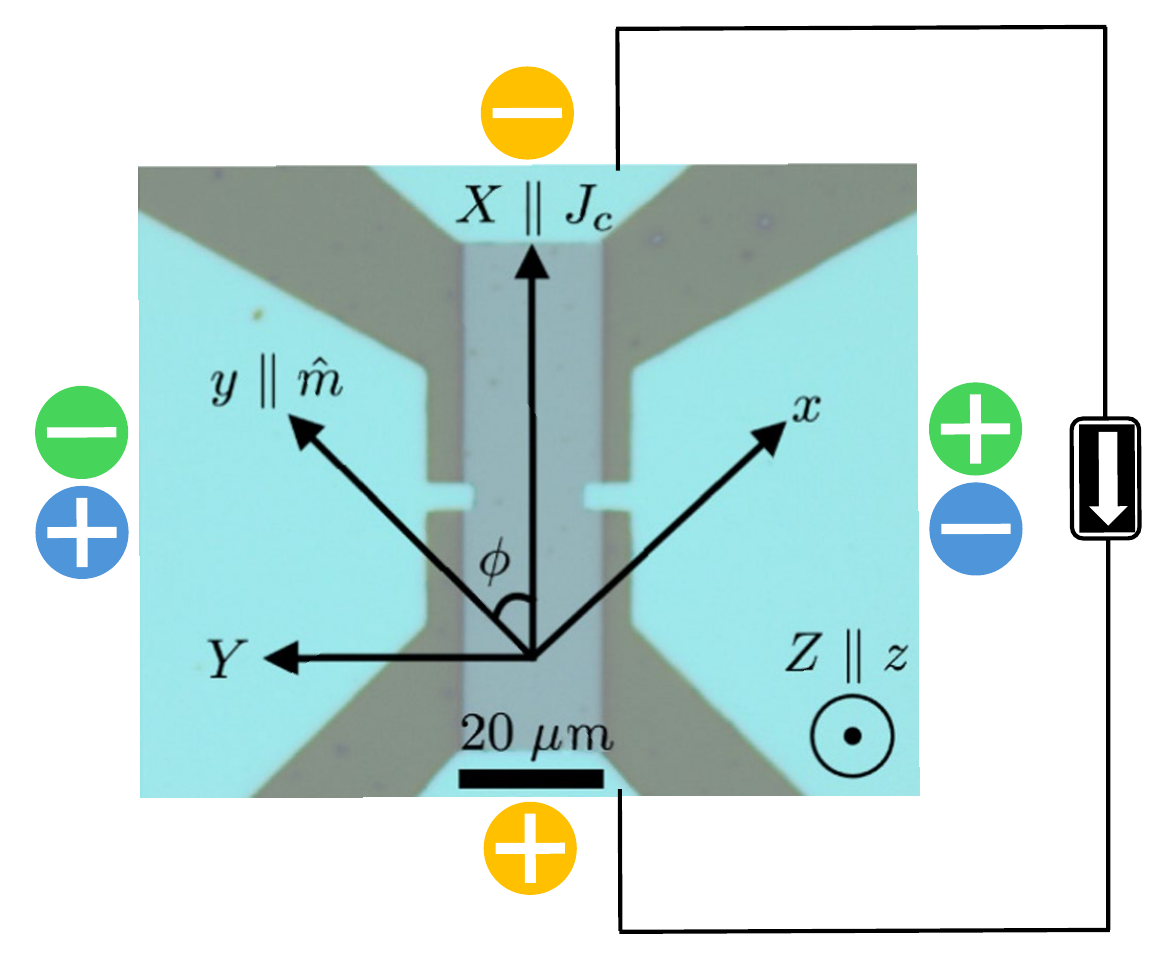}
    \caption{ST-FMR polarity conventions. Pictured is a top-down view of an ST-FMR spin-source/magnet bilayer device with the spin-source layer on the bottom, reproduced from reference \cite{karimeddiny2020transverse}. Given an applied current polarity (arrow), the colored pairs of $+$ and $-$ signs indicate different conventions for measuring voltages. }
    \label{fig: STFMRconvention}
\end{figure}

The probe polarity conventions used throughout this Perspective are implicit in the definitions of the magnetoresistances given in the main text. 
For longitudinal measurements, the expression $R_{\rm XX}=R_0+R_{\mathrm{AMR}} m_{\rm X}^2$,  along with the tacit convention that $R_0>0$, imply that the longitudinal sense leads are connected according to the orange polarity shown in Fig.~\ref{fig: STFMRconvention}. 
For the Hall measurement, in this paper we have chosen the convention $R_{\rm XY}=R_{\rm PHE}m_{\rm X}m_{\rm Y}+R_{\rm AHE}m_{\rm Z}$ with  $R_{\rm PHE}$ having the same sign as $R_{\rm AMR}$.  This corresponds to the green convention in Fig.~\ref{fig: STFMRconvention}.  The opposite convention (i.e., blue in Fig.~\ref{fig: STFMRconvention}) was used in ref.~\cite{karimeddiny2020transverse}.  
In this section, we discuss the reasons for our choice, and show how the fitting equations used for the artifact extraction depend on the convention used.\\

We adopt the green convention for four reasons: First, the PHE and AMR have the same sign, as mentioned above. That is, $R_{\rm PHE} = \left(\frac{W}{L}\right)R_{\rm AMR}$. Second, the (effective) PHE resistance is positive for both spin-Hall magneto-resistance and Fe, Co, Ni. Third, the AHE resistance sign matches that of the ordinary Hall effect for positive carriers. And fourth, the probe polarity with respect to the coordinate axes is consistent with the longitudinal probe.\\ 

Next, we discuss the impact these conventions have on the mixing voltages.  Taking into account, for simplicity, only the conventional components of current-induced torque, $\tau^0_\text{DL,Y}$ and $\gamma B^0_\text{FL,Y}$, along with the artifact signal, the longitudinal and lateral ST-FMR voltages have the form
\begin{equation}
\begin{aligned}
V_{\rm XX}=\sin(2\phi)\cos(\phi)\big[S^\text{AMR/art}_{\rm XX}S(B)+A^\text{AMR}_{\rm XX}A(B)\big]
\label{eq:stfmr_longV}
\end{aligned}
\end{equation}
\begin{equation}
\begin{aligned}
V_{\rm XY}=\cos(2\phi)\cos(\phi)\big[S^\text{PHE/art}_{\rm XY}S(B)+A^\text{PHE}_{\rm XY}A(B)\big] + \cos(\phi)\big[S^\text{AHE/art}_{XY}S(B)+A^\text{AHE}_{XY}A(B)\big]
\label{eq:stfmr_HallV}
\end{aligned}
\end{equation}
with the coefficients
$$
\begin{aligned}
S_{\rm X X}^{\mathrm{AMR}/\mathrm{art}} & =\frac{I_{\mathrm{rf}}}{2 \alpha \omega^{+}} R_{\mathrm{AMR}} \tau^0_\text{DL,Y}-\frac{L}{2} E_{\mathrm{art}}^0 \\
& \equiv S_{\rm X X}^{\mathrm{AMR}}+V_{\mathrm{art}}, \\
A_{\rm X X}^{\mathrm{AMR}} & =\frac{I_{\mathrm{rf}}}{2 \alpha \omega^{+}} R_{\mathrm{AMR}} \frac{\omega_2}{\omega} \gamma B^0_\text{FL,Y}, \\
S_{\rm X Y}^{\mathrm{PHE}/\mathrm{art}} & =-\frac{I_{\mathrm{rf}}}{2 \alpha \omega^{+}} R_{\mathrm{PHE}} \tau^0_\text{DL,Y}+\sigma\frac{W}{2} E_{\mathrm{art}}^0, \\
A_{\rm X Y}^{\mathrm{PHE}} & =-\frac{I_{\mathrm{rf}}}{2 \alpha \omega^{+}} R_{\mathrm{PHE}} \frac{\omega_2}{\omega} \gamma B^0_\text{FL,Y}, \\
S_{\rm X Y}^{\mathrm{AHE}/\mathrm{art}} & =\frac{I_{\mathrm{rf}}}{2 \alpha \omega^{+}} R_{\mathrm{AHE}} \gamma B^0_\text{FL,Y} +\sigma\frac{W}{2} E_{\mathrm{art}}^0, \\
A_{\rm X Y}^{\mathrm{AHE}} & =-\frac{I_{\mathrm{rf}}}{2 \alpha \omega^{+}} R_{\mathrm{AHE}} \frac{\omega_1}{\omega} \tau^0_\text{DL,Y}.
\end{aligned}
$$
The variable $\sigma$ in these expressions indicate which convention for the Hall voltage applies. 
Under the green convention $\sigma=1$, and for the blue convention $\sigma=- 1$. It is also important that the definitions for $R_{\rm PHE}$ and $R_{\rm AHE}$ used in these equations also differ in sign between conventions. The longitudinal convention is fixed to the orange polarity.\\

Notice that the only terms in these expressions that depend explicitly on $\sigma$ are the artifact terms contributing to the Hall measurements.  This is because both the measured Hall voltages (S and A coefficients) and the definitions of $R_\text{PHE}$ and $R_\text{AHE}$  switch sign upon reversing the Hall probes, while the artifact electric field $E^0_\text{art}$ is defined relative to fixed coordinate axes and does not change sign depending on the convention for measuring the Hall voltage. Therefore the terms describing the contributions to the Hall voltage from the artifact electric field must contain $\sigma$ so that they properly change sign upon a change in convention for the Hall voltage.  These equations with $\sigma = -1$ correspond to the result in ref.~\cite{karimeddiny2020transverse} for the blue convention in Fig.~\ref{fig: STFMRconvention}.\\

The artifact voltage can be extracted from the following  expressions for $\eta$, where $\eta$ is defined to be proportional to the ratio of the damping-like and field-like torques (see ref.~\cite{karimeddiny2020transverse})
\begin{align}
\eta=\frac{-A_{\rm X Y}^{\mathrm{AHE}}}{S_{\rm X Y}^{\mathrm{AHE}/\text{art}}+ (-\sigma)W\left(E_{\mathrm{art}}^0 / 2\right)}=\left\{\begin{array}{l}\frac{S_{\rm X Y}^{\mathrm{PHE}/\text{art}}+(-\sigma)W\left(E_{\mathrm{art}}^0 / 2\right)}{A_{\rm X Y}^{\mathrm{PHE}}} \\ \frac{S_{\rm X X}^{\mathrm{AMR}/\mathrm{art}}+L\left(E_{\mathrm{art}}^0/2\right)}{A_{\rm X X}^{\mathrm{AMR}}}\end{array}\right..
\label{eq:eta artifact}
\end{align}
The two expressions on the right-hand side are not independent, as they represent precisely the same physics because of the equivalency between the PHE and the AMR. Therefore the artifact voltage correction and measurement of $\tau^0_\text{DL,Y}$ can actually be performed without measuring the longitudinal mixing voltages. 
Equating the left-hand expression with the top right-hand expressions yields a solution for the artifact voltage from the transverse mixing signal:
\begin{equation}
E_\text{art}^0=\frac{
\sigma\left(S_{\rm X Y}^{\mathrm{PHE}/\mathrm{art}}+S_{\rm X Y}^{\mathrm{AHE}/\mathrm{art}}\right) \pm \sqrt{
\left(S_{\rm X Y}^{\mathrm{PHE}/\mathrm{art}} - S_{\rm X Y}^{\mathrm{AHE}/\mathrm{art}}\right)^2 -4 A_{\rm X Y}^{\mathrm{PHE}} A_{\rm X Y}^{\mathrm{AHE}}  }
}{W}.
\label{E1}
\end{equation}
If one performs both longitudinal and transverse ST-FMR, one can also equate the left-hand expression in Eq.~(\ref{eq:eta artifact}) with the bottom right-hand expression to solve for $E_{\rm art}^0$:
\begin{equation}
    E_{\rm art}^0=\frac{\left(S_{\rm X Y}^{\mathrm{AHE}/\mathrm{art}}L-\sigma S_{\rm X X}^{\mathrm{AMR}/\mathrm{art}}W\right) \pm \sqrt{
\left(S_{\rm X Y}^{\mathrm{AHE}/\mathrm{art}}L + \sigma S_{\rm X X}^{\mathrm{AMR}/\mathrm{art}}W\right)^2 +4\sigma WL A_{\rm X Y}^{\mathrm{AHE}} A_{\rm X X}^{\mathrm{AMR}}}}{\sigma WL}.
\label{E2}
\end{equation}
Equations (\ref{E1}) and (\ref{E2}) should agree modulo experimental noise unless there are reversed probe polarities or an error in setting a lock-in phase.\\

\end{document}